\documentclass[aps,prd,preprint,eqsecnum,tightenlines,nofootinbib,showpacs]{revtex4}
\usepackage[dvips]{graphicx}
\usepackage{epsf,subfigure}
\usepackage{amsfonts,amsmath}
\usepackage{relsize}
\usepackage{multirow}
\usepackage{array}

\newif\ifdraft
\drafttrue
\newif\ifpreprint
\preprinttrue
\newcommand{\subV}{\mathsmaller{\rm{V}}}
\newcommand{\subH}{\mathsmaller{\rm{H}}}
\newcolumntype{C}[1]{>{\centering\let\newline\\\arraybackslash\hspace{0pt}}m{#1}}

\def\fig#1{Fig.~{\ref{#1}}}

\def\eqn#1{Eq.~({\ref{#1}})}
\def\eqns#1#2{Eqs.~({\ref{#1}}) and~({\ref{#2}})} 
\def\sect#1{Section~{\ref{#1}}}
\def\sects#1#2{Sections~\ref{#1} and~\ref{#2}}

\def\tab#1{Table~{\ref{#1}}}

\def\App#1{Appendix~{\ref{#1}}}

\def\nn{\nonumber}
\def\NeqFour{\mathcal{N}=4}

\def\NeqEight{\mathcal{N}=8}

\def\tree{{\rm tree}}
\def\oneloop{{(1)}}
\def\twoloop{{(2)}}

\def\n{\tilde n}
\def\f{\tilde f}

\def\eps{\epsilon}

\def\tree{{\rm tree}}

\def\Tr{{\rm Tr}}

\def\spa#1.#2{\left\langle#1\,#2\right\rangle}
\def\spb#1.#2{\left[#1\,#2\right]}

\newbox\charbox
\newbox\slabox
\def\s#1{{      % Feynman slash
        \setbox\charbox=\hbox{$#1$}
        \setbox\slabox=\hbox{$/$}
        \dimen\charbox=\ht\slabox
        \advance\dimen\charbox by -\dp\slabox
        \advance\dimen\charbox by -\ht\charbox
        \advance\dimen\charbox by \dp\charbox
        \divide\dimen\charbox by 2
        \raise-\dimen\charbox\hbox to \wd\charbox{\hss/\hss}
        \llap{$#1$}
}}

\begin{document}

\title{
\ifpreprint
 \hbox{\rm \normalsize  UCLA/13/TEP/105} 
\hbox{$\null$}
\fi
\Large The Ultraviolet Structure of Half-Maximal Supergravity
with Matter Multiplets at Two and Three Loops
}
 
\author{Zvi~Bern$^a$, Scott~Davies$^a$ and Tristan~Dennen$^{b}$}

\affiliation{
$a$ Department of Physics and Astronomy, University of California 
at Los Angeles\\ 
 Los Angeles, CA 90095-1547, USA \\ 
$\null$ \\
$b$ Niels Bohr International Academy and Discovery Center\\
The Niels Bohr Institute\\
Blegdamsvej 17, DK-2100 Copenhagen, Denmark\\
}

\vskip .5 cm
\begin{abstract}
Using the duality between color and kinematics, we construct the two-
and three-loop amplitudes of half-maximal supergravity with matter multiplets and
show that new divergences occur in $D=4$ and $D=5$.  Bossard, Howe and
Stelle have recently conjectured the existence of 16-supercharge
off-shell harmonic superspaces in order to explain the ultraviolet
finiteness of pure half-maximal supergravity with no matter multiplets
in $D=4$ at three loops and in $D=5$ at two loops.  By assuming the
required superspace exists in $D=5$, they argued that no new
divergences should occur at two loops even with the addition of
abelian-vector matter multiplets.  Up to possible issues with the
$SL(2,\mathbb{R})$ global anomaly of the theory, they reached a
similar conclusion in $D=4$ for two and three loops.  The divergences we find
contradict these predictions based on the existence of the desired off-shell superspaces.
Furthermore, our
$D=4$ results are incompatible with the new divergences being due to
the anomaly.  We find that
the two-loop divergences of half-maximal supergravity are directly
controlled by the divergences appearing in ordinary nonsupersymmetric
Yang-Mills theory coupled to scalars, explaining why half-maximal
supergravity develops new divergences when matter multiplets are
added.  We also provide a list of one- and two-loop counterterms that
should be helpful for constraining any future potential explanations
of the observed vanishings of divergences in pure half-maximal
supergravity.
\end{abstract}

\pacs{04.65.+e, 11.15.Bt, 11.30.Pb, 11.55.Bq \hspace{1cm}}

\maketitle

\section{Introduction}

The possibility of finding ultraviolet finite supergravity
theories~\cite{Supergravity} has been reopened in recent years due to
the discovery of new unexpected ultraviolet cancellations.  Such
theories were intensely studied in the late 1970's and early 1980's as
possible fundamental theories of gravity, but fell out of favor when
the way forward seemed blocked by the likely appearance of
nonrenormalizable ultraviolet divergences. (For a review article from that era
see, for example, Ref.~\cite{HoweStelleReview}.)  At the time it was not
possible to definitively determine the divergence structure of
supergravity theories because there were no means available for
carrying out the required computations.  Today thanks to the unitarity
method~\cite{UnitarityMethod} and the recently uncovered duality
between color and kinematics~\cite{BCJ,BCJLoop}, we have the ability
to address this.

Explicit calculations~\cite{GravityThree,GravityFour} show that
$\NeqEight$ supergravity~\cite{N8Sugra} is finite for dimensions $d <
6/L + 4$, at least through $L=4$ loops.  If one were to extrapolate
the observed cancellations, assuming no new ones occur, simple power
counting suggests that no divergence can occur in the theory prior to
seven loops.  Indeed, new detailed studies of the known standard symmetries of
$\NeqEight$ supergravity demonstrate that no valid counterterms can be
found prior to the seventh loop order, but at seven loops a $D^8 R^4$
counterterm can be constructed that appears to obey all known
symmetries~\cite{SevenLoopGravity}.  An explicit expression for the
potential counterterm was written down in Ref.~\cite{VanishingVolume}.
These facts suggest that $\NeqEight$ supergravity diverges at seven
loops; of course, this assumes that all symmetries and structures have
been properly taken into account.

When similar arguments are applied to pure half-maximal
supergravity~\cite{N4Sugra} at three loops in $D=4$ and two loops in
$D=5$, counterterms valid under all known symmetries have been
found~\cite{VanishingVolume}.  However, we now know
from explicit computations that there are no divergences corresponding
to these counterterms~\cite{N4gravThreeLoops,HalfMax5D}. In addition,
arguments for finiteness in these cases based on string theory have
been given in Ref.~\cite{VanhoveN4}.  At three loops in $D=4$, there is
only one available counterterm in pure $\NeqFour$
supergravity~\cite{VanishingVolume,BossardHoweStelle5D,BHSNew}, so the
fact that its coefficient vanishes implies that the full theory is
three-loop finite.

The surprisingly good ultraviolet behavior of pure half-maximal
supergravity has led to conjectures to explain its origin.  One
conjecture is that it is due to a hidden superconformal
symmetry~\cite{RenataSergioN4}.  A more controversial conjecture is
that the potential counterterms break relevant duality symmetries
modified by quantum corrections~\cite{KalloshN4}.  A third conjecture
is that the duality between color and kinematics leads to cancellation
of the ultraviolet infinities in ${\cal N} \ge 4$ four-dimensional
supergravity by the same mechanism that prevents forbidden loop-level
color tensors from appearing in pure nonsupersymmetric Yang-Mills
divergences~\cite{HalfMax5D}.  

On the other hand, Bossard, Howe and Stelle have given a potential symmetry
explanation that would not require any new ``miracles'' beyond those
of supersymmetry and ordinary duality symmetries.  By conjecturing the
existence of appropriate harmonic superspaces in $D=4$ and $D=5$
manifesting all 16 supercharges off
shell~\cite{BossardHoweStelle5D,BHSNew}, they have explained the
observed ultraviolet cancellations.  If true, it would predict that
ultraviolet divergences start at four loops in both $D=4$ and $D=5$ in
pure half-maximal supergravity.  (An artifact of dimensional
regularization is that there are no three-loop divergences in $D=5$.)
While it is unclear how to construct the conjectured superspaces, one
can still deduce consequences by assuming their existence. Following
this reasoning, Bossard, Howe and Stelle have shown~\cite{BHSNew} that
no new divergences should appear at two loops in $D=5$ even after
adding matter multiplets~\cite{deRoo,Awada}, were the desired
16-supercharge superspace to exist.  In $D=4$ the situation is
similar, leading us to the issue of whether the anomaly in the rigid
$SL(2,\mathbb{R})$ duality symmetry~\cite{MarcusAnomaly} might play a
role in the appearance of new divergences in matter-multiplet
amplitudes~\cite{BHSNew}.  (We note that the study of matter
multiplets in supergravity theories and their divergence properties
has a long history~\cite{SugraMatter}.)

The predictions of Ref.~\cite{BHSNew} motivated us to compute the
coefficients of two-loop four-point divergences in half-maximal
supergravity including vector multiplets in $D=4,5,6$ to definitively
demonstrate that there are nonvanishing divergences in all these
dimensions, as well as to give their precise form.  Our two-loop $D=5$
result is in direct conflict with the predicted
finiteness~\cite{BHSNew} based on assuming the existence of an
off-shell 16-supercharge superspace.  We also find a two-loop
divergence in $D=4$ after subtracting the one-loop subdivergences.
This two-loop divergence happens to resemble an iteration of the
one-loop divergence, so to remove any potential doubts as to whether
there are new divergences, we also calculate the complete set of
three-loop divergences in $D=4$, making it clear that there are indeed new
divergences.  Furthermore, we find that the explicit two- and
three-loop $D=4$ results are not of the form required had they been
due to the anomaly. From Ref.~\cite{BHSNew}, 
it therefore appears that the desired 16-supercharge
superspaces exist in neither $D=4$ nor $D=5$.

As discussed in Ref.~\cite{HalfMax5D}, for one and two loops 
in half-maximal supergravity, whenever
divergences in the coefficients of certain color tensors are forbidden
in gauge theory, divergences also cancel from the corresponding
half-maximal supergravity amplitudes.  In particular, the lack of one-loop divergences in dimensions $D<8$ in
half-maximal supergravity amplitudes with four external states of the
graviton multiplet is a direct consequence of the lack of divergences
in those terms in gluon amplitudes proportional to the independent
one-loop color tensor.  Since the four-scalar amplitude of Yang-Mills
theory coupled to scalars does contain divergences in terms containing
the one-loop color tensor even in $D=4$, the corresponding four-matter
multiplet amplitude of $\NeqFour$ supergravity with matter multiplets
also diverges.  This is in agreement with the result found long ago by
Fischler~\cite{Fischler} and Fradkin and Tseytlin~\cite{Arkady}.  At
two loops the situation is similar. The four-point amplitudes with
pure external graviton multiplet states are ultraviolet finite in
$D=4,5$ because all corresponding gauge-theory divergences contain
only tree-level color tensors.  However, because four-scalar
amplitudes of gauge theory, both in $D=4$ and $D=5$, contain
divergences in the coefficients of the independent two-loop color
tensors, corresponding two-loop four-matter multiplet amplitudes of
half-maximal supergravity must also diverge.

At three and higher loops, the situation is more complicated because
loop momenta appear in the maximal super-Yang-Mills duality-satisfying
numerators, so the supergravity integrals are no longer the same ones
as those appearing in gauge theory.  Because of this, a link between the
divergences of half-maximal supergravity and those of
nonsupersymmetric gauge theory will require nontrivial
integral identities and remains speculative~\cite{HalfMax5D}.

To carry out our investigation, we construct half-maximal supergravity
amplitudes via the duality between color and
kinematics~\cite{BCJ,BCJLoop}.  In this way, gravity loop integrands
are obtained from a pair of corresponding gauge-theory loop
integrands. The key to this construction is to find a representation
where one of the two gauge-theory amplitudes manifestly exhibits the
duality between color and kinematics.  Here we obtain half-maximal
supergravity with matter multiplets from a direct product of maximal
super-Yang-Mills theory and nonsupersymmetric Yang-Mills theory with
interacting scalars.  The required two-loop super-Yang-Mills amplitude
in a form where the duality is manifest was given long ago in
Refs.~\cite{BRY,BDDPR}, while the desired form of the 
three-loop amplitude was given more
recently in Ref.~\cite{BCJLoop}.  The nonsupersymmetric Yang-Mills
theory coupled to $n_{\subV}$ scalars is conveniently obtained by
dimensionally reducing pure Yang-Mills theory from $D + n_{\subV}$
dimensions to $D$ dimensions, matching the construction of
half-maximal supergravity with $n_{\subV}$ matter multiplets by
dimensional reduction of pure half-maximal
supergravity~\cite{RaduMatter}.

Once we have the integrands for the amplitudes, we need to extract the
ultraviolet singularities. The basic procedure for doing so has been
long understood~\cite{MarcusSagnotti} and has been applied recently to
a variety of supergravity and super-Yang-Mills
calculations~\cite{GravityFour,ck4l,N4gravThreeLoops,HalfMax5D}.  Here
we will explain in some detail the procedure that we use to extract
ultraviolet divergences in the presence of integral-by-integral
subdivergences.  This procedure was already used in
Ref.~\cite{N4gravThreeLoops} to demonstrate the ultraviolet finiteness
of all three-loop four-point amplitudes of pure $\NeqFour$
supergravity in four dimensions.

This paper is organized as follows.  In \sect{ReviewSection}, we
briefly review some basic facts of the duality between color and
kinematics and the double-copy construction of gravity.  We also
explain the structure of one- and two-loop four-point amplitudes in
half-maximal supergravity with $n_{\subV}$ abelian matter multiplets.
Then in \sect{ProcedureSection} we describe the construction of the
integrand and the integration methods used to extract the ultraviolet
divergences.  We give our results for the one-loop and two-loop
divergences of half-maximal supergravity with matter multiplets in
\sects{OneLoopDivergenceSection}{TwoLoopDivergenceSection}, with
\sect{TwoLoopDivergenceSection} also containing our $D=4$ three-loop
results.  Finally we present our conclusions and outlook in
\sect{ConclusionSection}.  An appendix listing valid counterterms as
well as their numerical coefficients is also given.

%%%%%%%%%%%%%%%%%%%%%%%%%%%%%%%
\section{Basic Setup}
\label{ReviewSection}

The duality between color and kinematics and the associated gravity
double-copy property~\cite{BCJ,BCJLoop} make it straightforward to
construct supergravity amplitudes once corresponding gauge-theory
amplitudes are arranged into a form that makes the duality manifest.
(For a recent review of this duality and its application, see
Ref.~\cite{HenrikJJReview}.)  While the duality remains a conjecture
at loop level, we will use it only for one-, two- and three-loop
four-point amplitudes where it is known to hold in maximally
supersymmetric Yang-Mills theory.  We use it to map out the two- and
three-loop divergence structure of half-maximal supergravity with
$n_{\subV}$ abelian matter multiplets.  We start by first giving a
brief summary of the duality before giving a number of formulas that
are useful for one- and two-loop amplitudes in half-maximal
supergravity~\cite{OneLoopN4,TwoLoopN4,HalfMax5D}.

\subsection{Duality between color and kinematics}

The gauge-theory duality between color and kinematics is conveniently
described in terms of graphs with only cubic vertices. 
Using such graphs, any
$m$-point $L$-loop gauge-theory amplitude with all particles in the
color adjoint representation can be written as
\begin{equation}
 {\cal A}^{L-\rm loop}_m =  {i^L} {g^{m-2 +2L }}
\sum_{{\cal S}_m} \sum_{j}{\int \prod_{l=1}^L \frac{d^{d} p_l}{ (2 \pi)^{d}}
  \frac{1}{S_j}  \frac {n_j c_j}{\prod_{\alpha_j}{p^2_{\alpha_j}}}}\,,
\label{LoopGauge} 
\end{equation}
where the sum labeled by $j$ runs over the set of distinct
non-isomorphic graphs.  Any contact terms in the amplitude can be
expressed in terms of graphs with only cubic vertices by multiplying
and dividing by appropriate propagators.  The product in the
denominator runs over all Feynman propagators of graph $j$.  The
integrals are over $L$ independent $d$-dimensional loop momenta.  The
symmetry factor $S_j$ removes over counts from the sum over
permutations of external legs indicated by ${\cal S}_m$ and from
internal symmetry factors. The $c_j$ are color factors obtained by
dressing every three-vertex with a group-theory structure constant,
\begin{equation}
\f^{abc} = i \sqrt{2} f^{abc}=\Tr([T^{a},T^{b}]T^{c}) \,,
\label{fabcdef}
\end{equation}
and $n_j$ are kinematic numerators of graph $j$ depending on
momenta, polarizations and spinors.  If a superspace formulation
is used, $n_j$ can also depend on Grassmann parameters. 

%%%%%%%%%%%%%% FIGURE %%%%%%%%%%%%%
\begin{figure}
\includegraphics[scale=.45]{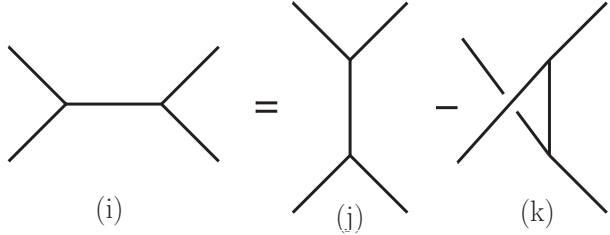}
\caption{The basic Jacobi relation for either color or numerator
  factors given in \eqn{BCJDuality}.  These three diagrams can be
  embedded in a larger diagram, including loops. }
\label{BCJFigure}
\end{figure}
%%%%%%%%%%%%%%%%%%%%%%%%%%%

The conjectured duality of Refs.~\cite{BCJ,BCJLoop} states that to all
loop orders, there exists a form of (super-)Yang-Mills amplitudes
where kinematic numerators satisfy the same algebraic relations as
color factors.  In these theories, this amounts to imposing the same
Jacobi identities on the kinematic numerators as satisfied by adjoint-representation color factors,
\begin{equation}
c_i = c_j - c_k \;  \Rightarrow \;  n_i = n_j - n_k \,,
\label{BCJDuality}
\end{equation}
where the indices $i,j,k$ denote the diagram to which the color
factors and numerators belong.  The basic Jacobi identity is
illustrated in \fig{BCJFigure} and can be embedded in arbitrary
diagrams.  The numerator factors are also required to have the same
antisymmetry properties as color factors.  In general, the duality
relations (\ref{BCJDuality}) work only after appropriate nontrivial
rearrangements of the amplitudes.

When a representation of an amplitude is found where the duality
(\ref{BCJDuality}) is made manifest, we can obtain corresponding
gravity loop integrands simply by replacing color factors in a 
gauge-theory amplitude by kinematic numerators of a second gauge-theory amplitude.  This gives the double-copy form of
corresponding gravity amplitudes~\cite{BCJ,BCJLoop},
\begin{equation}
{\cal M}^{L-\rm loop}_m =  {i^{L+1}} {\Bigl(\frac{\kappa}{2}\Bigr)^{m-2+2L}} \,
\sum_{{\cal S}_m} \sum_{j} {\int \prod_{l=1}^L \frac{d^{d} p_l}{(2 \pi)^{d}}
 \frac{1}{S_j} \frac{n_j \n_j}{\prod_{\alpha_j}{p^2_{\alpha_j}}}} \,.
\hskip .7 cm 
\label{DoubleCopy}
\end{equation}
Generalized gauge invariance implies that only one of the two sets of
numerators $n_j$ or $\n_j$ needs to satisfy the duality relation
(\ref{BCJDuality})~\cite{BCJLoop,BCJSquare}.  At tree level, the
double-copy formula (\ref{DoubleCopy}) encodes the Kawai-Lewellen-Tye
(KLT)~\cite{KLT} relations between gravity and gauge-theory
amplitudes~\cite{BCJ}.

In this paper, we will construct amplitudes for half-maximal
supergravity with matter multiplets as a double copy of maximally
supersymmetric Yang-Mills amplitudes and nonsupersymmetric
Yang-Mills amplitudes coupled to interacting scalars.  The desired
scalars arise from dimensional reduction of pure Yang-Mills theory.
For such scalars the conjectured duality holds automatically when it
holds in higher-dimensional pure Yang-Mills theory.  We will not need
duality-satisfying representations of the nonsupersymmetric
amplitudes, given that we have them on the maximal super-Yang-Mills
side.

%%%%%%%%%%%%%%%%%%%%%%%

\subsection{Amplitude relations at one and two loops}
\label{AmplitudesSubsection}

As explained in Refs.~\cite{OneLoopN4,TwoLoopN4,HalfMax5D}, the one-
and two-loop four-point amplitudes of pure half-maximal supergravity
are easily obtained from corresponding amplitudes in nonsupersymmetric
gauge theory.  Here we extend this slightly by noting that the same
holds for half-maximal supergravity amplitudes including
abelian-vector multiplets.

%%%%%%%%%%%%%%%%%%%%
%FIGURE
%
\begin{figure}
\begin{center}
\includegraphics[scale=0.6]{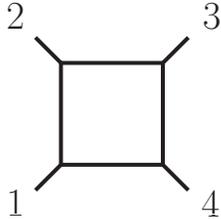}
\caption{
The one-loop box diagram.  The  one-loop color factor $c^{\oneloop}_{1234}$
is obtained by dressing each vertex with an $\f^{abc}$. }
\label{BoxFigure}
\end{center}
\end{figure}
%%%%%%%%%%%%%%%%%%%%%%

The double-copy construction of a half-maximal supergravity amplitude
starts by writing the corresponding nonsupersymmetric gauge-theory
amplitude in a convenient color decomposition, then replacing color
factors by super-Yang-Mills numerators that satisfy the duality
between color and kinematics.  A color-dressed four-point one-loop
gauge-theory amplitude with all particles in the adjoint
representation can be expressed as~\cite{DixonMaltoniColor}
\begin{eqnarray}
\mathcal{A}^{\oneloop}(1,2,3,4)=g^4\Bigl[
  c^{\oneloop}_{1234}A^{\oneloop}(1,2,3,4)
+ c^{\oneloop}_{1342}A^{\oneloop}(1,3,4,2)
+ c^{\oneloop}_{1423}A^{\oneloop}(1,4,2,3)\Bigr]\,.
\label{OneLoopFourPtGauge}
\end{eqnarray} 
The $c^{\oneloop}_{1234}$ are the color factors of a box diagram,
illustrated in \fig{BoxFigure}, with consecutive external legs
$(1,2,3,4)$ and with vertices dressed with structure constants $\f^{abc}$,
defined in \eqn{fabcdef}. The $A^{\oneloop}$ are one-loop color-ordered
amplitudes~\cite{ColorLoop}. This color decomposition holds just as
well whether the external particles are adjoint scalars or gluons and
does not depend on supersymmetry.

To obtain the one-loop half-maximal supergravity amplitudes, we simply
replace the gauge coupling with the gravitational one and the color factors in \eqn{OneLoopFourPtGauge} with maximal
super-Yang-Mills duality-satisfying kinematic
numerators~\cite{OneLoopN4}, \begin{equation}
c_{ijkl}^{\oneloop} \rightarrow n_{ijkl}^{\oneloop} \,, 
 \hskip 2 cm g^4 \rightarrow i\Bigl(\frac{\kappa}{2}\Bigr)^4\,,
\end{equation}
where~\cite{GSB}
\begin{equation}
n_{1234}^{\oneloop} = n_{1342}^{\oneloop} = n_{1423}^{\oneloop} = s t A^\tree_{Q=16} (1,2,3,4)\,,
\end{equation}
and $A^{\tree}_{Q=16}(1,2,3,4)$ is the four-point tree amplitude of
maximal 16-supercharge super-Yang-Mills theory, valid for all states
of the theory.  (See for example
Eq.~(2.8) of Ref.~\cite{SuperSum} for the explicit form of these
tree amplitudes in $D=4$.)
This gives us a rather simple formula for one-loop
four-point amplitudes in half-maximal supergravity~\cite{OneLoopN4},
\begin{equation}
\mathcal{M}^{\oneloop}_{Q=16}=
i\Bigl(\frac{\kappa}{2}\Bigr)^4 s t A^{\rm tree}_{Q=16}(1,2,3,4)
\Big[A^{\oneloop}(1,2,3,4) + A^{\oneloop}(1,3,4,2) + A^{\oneloop}(1,4,2,3) \Big]\,.
\hskip .5 cm 
\label{FourPointSupergravity}
\end{equation} 
This formula is valid for all matter- and graviton-multiplet states of
half-maximal supergravity.  This simple replacement rule means that
the supergravity divergences can be read off directly from the
gauge-theory divergences. In particular, we can read off the
divergences of half-maximal supergravity with $n_{\subV}$ vector
multiplets directly from the corresponding divergences of
nonsupersymmetric Yang-Mills theory coupled to $n_{\subV}$ scalars.

The expression (\ref{FourPointSupergravity}) automatically satisfies
the unitarity cuts if the input gauge-theory amplitudes are correct.
This is because once the maximally supersymmetric numerators that
satisfy the duality between color and kinematics are used, the cuts
necessarily match those obtained by feeding in gravity tree amplitudes
obtained by either the double-copy formula or the KLT relations.  In
addition, this formula has been used~\cite{OneLoopN4} to reproduce
known expressions~\cite{DunbarNorridge} for the integrated amplitudes
in ${\cal N} = 4,6$ supergravity, when $A^{\oneloop}$ is taken to represent
more general, possibly supersymmetric, gauge-theory amplitudes.  It
also matches the known expression for $\NeqEight$
supergravity~\cite{BDDPR}.

As explained in Ref.~\cite{HalfMax5D}, we can line up the divergences of
supergravity with those appearing in the independent one-loop 
color tensor of the color basis given in Appendix B of Ref.~\cite{Neq44np}
(see also Ref.~\cite{Naculich}).
In this color basis we have
\begin{equation}
b^{\oneloop}_1 \equiv c_{1234}^{\oneloop} \,, \hskip 2 cm 
c^{\oneloop}_{1342} = b^{\oneloop}_1 + \cdots \,,  \hskip 2 cm 
c^{\oneloop}_{1423} = b^{\oneloop}_1 + \cdots \,,
\label{OneLoopColorBasis}
\end{equation}
where ``$\null +\cdots$'' represents dropped terms proportional to
the tree-level color tensors,
\begin{equation}
b^{(0)}_1  = \f^{a_1 a_2 b} \f^{b a_3 a_4} \,, \hskip 2 cm 
b^{(0)}_2  =  \f^{a_2 a_3 b} \f^{b a_4 a_1} \,.
\label{TreeColorTensor}
\end{equation}
 After expressing all the color
factors in the basis (\ref{OneLoopColorBasis}), the gauge-theory amplitude (\ref{OneLoopFourPtGauge})
can be expressed as
\begin{equation}
\mathcal{A}^{\oneloop}(1,2,3,4) =g^4 b^{\oneloop}_1 \Bigl(A^{\oneloop}(1,2,3,4)
         + A^{\oneloop}(1,3,4,2) + A^{\oneloop}(1,4,2,3) \Bigr) + \cdots \,. \hskip .3 cm 
\label{OneLoopColorBasisAmplitude}
\end{equation}
In this form, we see that the 
half-maximal supergravity amplitudes line up with those 
terms in the nonsupersymmetric gauge-theory amplitudes 
containing the independent one-loop color tensor.

This remarkable relation between the one-loop half-maximal
supergravity amplitudes (\ref{FourPointSupergravity}) and the parts of
gauge-theory amplitudes containing the independent one-loop color tensor
(\ref{OneLoopColorBasisAmplitude}) allows us to obtain the supergravity
amplitude simply by converting to a color basis, dropping the
tree-level color factors, and then replacing the one-loop color
tensor and gauge coupling,
\begin{equation}
 b^{\oneloop}_1 \rightarrow s t A^{\tree}_{Q=16}(1,2,3,4)\,, \hskip 2 cm 
 g^4 \rightarrow i\Bigl(\frac{\kappa}{2}\Bigr)^4\,.
\label{N4Replacement}
\end{equation}
This works as well at the integrated level, so that once we have the
gauge-theory divergences, the substitution (\ref{N4Replacement})
directly gives us the corresponding divergences in one-loop
half-maximal supergravity with or without matter multiplets.

%%%%%%%%%%%%%%%%%%%%
%FIGURE
%
\begin{figure}
\begin{center}
\includegraphics[scale=0.6]{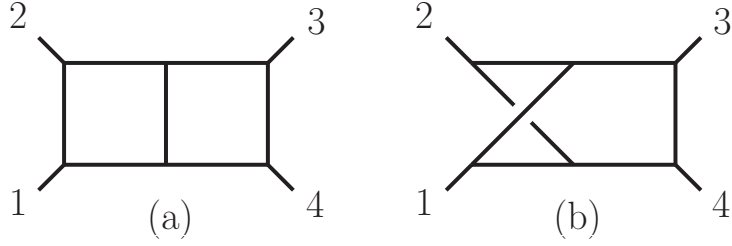}
\caption{The planar and nonplanar double-box graphs.  The $c^{\rm
    P}_{1234}$ and $c^{\rm NP}_{1234}$ color factors are obtained by
  dressing each vertex with an $\f^{abc}$.  }
\label{DoubleBoxFigure}
\end{center}
\end{figure}
%%%%%%%%%%%%%%%%%%%%%%

The situation is similar for two-loop four-point amplitudes.  At two
loops any color-dressed gauge-theory amplitude with only
adjoint-representation particles can be conveniently written
as~\cite{DixonMaltoniColor,OneLoopN4,TwoLoopN4}
\begin{eqnarray}
\mathcal{A}^{\twoloop}(1,2,3,4) &=&
g^6 \hskip -0.3 cm \sum_{x \in \{\rm P, NP\}} \Big[ c^{x}_{1234} A^{x}(1,2,3,4)
+ c^{x}_{3421} A^{x}(3,4,2,1) \label{YMTwoloop} 
+ c^{x}_{1423} A^{x}(1,4,2,3) \nn\\
 && \hskip .5cm   \null 
+ c^{x}_{2341} A^{x}(2,3,4,1) 
+ c^{x}_{1342} A^{x}(1,3,4,2) + c^{x}_{4231} A^{x}(4,2,3,1) 
\Big]\,, \hskip 1 cm 
\end{eqnarray} 
where the sum runs over the planar and nonplanar contributions.  Here
$c^{\rm P}_{1234}$ and $c^{\rm NP}_{1234}$ are the color factors
obtained by dressing the planar and nonplanar double-box graphs in
\fig{DoubleBoxFigure} with structure constants $\f^{abc}$, defined in 
\eqn{fabcdef}. The $A^{\rm
  P}$ and $A^{\rm NP}$ are planar and nonplanar
partial amplitudes.  

To obtain supergravity amplitudes, we replace the gauge coupling with
the gravitational one and the color factors in \eqn{YMTwoloop} with
the super-Yang-Mills numerators,
\begin{equation}
c^{\rm P}_{ijkl} \rightarrow n^{\rm P}_{ijkl}\,, \hskip 2cm 
c^{\rm NP}_{ijkl} \rightarrow n^{\rm NP}_{ijkl}\,,
 \hskip 2 cm g^6 \rightarrow i\Bigl(\frac{\kappa}{2}\Bigr)^6\,.
\end{equation}
where~\cite{BRY,BDDPR} 
\begin{eqnarray}
&&  n^{x}_{1234} = s K \,, \hskip 1 cm 
 n^{x}_{3421} = s K \,, \hskip 1 cm 
n^{x}_{1423} = t K \,, \nn \\ 
&& n^{x}_{2341} = t K \,,  \hskip 1 cm 
 n^{x}_{1342} = u K \,, \hskip 1 cm 
 n^{x}_{4231} = u K\,, 
\end{eqnarray}
and $x\in {\rm P,NP}$.  The factor $K$ is the fully crossing-symmetric
prefactor,
\begin{equation}
K =  s t A^\tree_{Q=16}(1,2,3,4) \,.
\end{equation}
In this way we immediately obtain the four-point two-loop
amplitude of half-maximal supergravity~\cite{OneLoopN4,TwoLoopN4},
\begin{eqnarray}
\mathcal{M}^{\twoloop}_{Q=16}(1,2,3,4) &=&
i\Bigl(\frac{\kappa}{2}\Bigr)^6 s t A^{\rm tree}_{Q=16}(1,2,3,4)
 \hskip -0.3 cm \sum_{x \in \{\rm P, NP\}} \Bigl[
 s (A^{x}(1,2,3,4) + A^{x}(3,4,2,1)  ) \nonumber\\
&& \hskip .2 cm
+ t ( A^{x}(1,4,2,3)  + A^{x}(2,3,4,1))
+ u ( A^{x}(1,3,4,2)  + A^{x}(4,2,3,1)) \Bigr]\,. \nonumber\\ 
\label{TwoLoopSugraBCJ}
\end{eqnarray}
As for one loop, this holds for all states of the graviton or vector
multiplets of half-maximal supergravity.  A nontrivial check that has
been carried out on this formula~\cite{TwoLoopN4} is that when the appropriate
integrated gauge-theory amplitudes~\cite{TwoLoopsYM} are inserted, it correctly
reproduces the known infrared singularities of ${\cal N} \ge 4$ supergravity
theories~\cite{SchnitzerIR}.

We can line up the supergravity amplitude with the contributions
proportional to the two independent two-loop color tensors~\cite{HalfMax5D},
\begin{eqnarray}
 b^{\twoloop}_1 &\equiv& c^{\rm P}_{1234}  \,, \hskip 1.5 cm 
 b^{\twoloop}_2 \equiv c^{\rm P}_{2341} \,, 
\end{eqnarray}
where the other color factors in the amplitude (\ref{YMTwoloop}) can
be expressed in terms of these:
\begin{eqnarray}
&& c^{\rm P}_{3421} = b^{\twoloop}_1 + \cdots \,, \hskip 3 cm 
c^{\rm P}_{1423} = b^{\twoloop}_2 + \cdots \,, \nn \\
&& 
c^{\rm P}_{1342} = -b^{\twoloop}_1 - b^{\twoloop}_2 +\cdots \,, \hskip 1.6 cm 
c^{\rm P}_{4231} = -b^{\twoloop}_1 - b^{\twoloop}_2 + \cdots\,,
\label{TwoLoopColorBasis}
\end{eqnarray}
and the ``$\null+ \cdots$'' represents dropped terms containing
lower-loop color tensors.  
The nonplanar color factors are the same as the planar ones, 
up to corrections proportional to lower-loop color tensors:
\begin{equation}
c^{\rm NP}_{ijkl} = c^{\rm P}_{ijkl} + \cdots \,.
\end{equation}
Substituting these into \eqn{YMTwoloop} gives 
\begin{eqnarray}
\mathcal{A}^{\twoloop}(1,2,3,4) &=&
g^6 \hskip -0.3 cm \sum_{x \in \{\rm P, NP\}} \Bigl[
 b_1^{\twoloop} (A^{x}(1,2,3,4) + A^{x}(3,4,2,1) - 
          A^{x}(1,3,4,2) - A^{x}(4,2,3,1) ) \nonumber\\
&& \hskip 1.5 cm \null
+ b_2^{\twoloop} ( A^{x}(1,4,2,3)  + A^{x}(2,3,4,1)
 - A^{x}(1,3,4,2)  - A^{x}(4,2,3,1) \Bigr] \nonumber\\ 
&& \hskip 2 cm \null
+ \cdots\,.
\label{TwoLoopGaugeColorBasis}
\end{eqnarray}
This lines up with the supergravity
expression~\eqref{TwoLoopSugraBCJ} once we replace $u = -s-t$.  Comparing
\eqn{TwoLoopGaugeColorBasis} to the two-loop supergravity 
expression shows that we can obtain half-maximal supergravity
divergences directly from the gauge-theory ones by going to 
the color basis (\ref{TwoLoopColorBasis}), dropping all one-loop
and tree color tensors, and then replacing
\begin{equation}
b_1^{\twoloop} \rightarrow s^2 t A_{Q=16}^\tree(1,2,3,4) \,, \hskip 1 cm 
b_2^{\twoloop} \rightarrow s t^2 A_{Q=16}^\tree(1,2,3,4) \,, \hskip 1 cm 
g^6 \rightarrow i\Bigl(\frac{\kappa}{2}\Bigr)^6\,.
\label{TwoLoopSubstitution}
\end{equation}

We note that in general at higher loops, one should not use a color
basis to make numerator substitutions because it assumes that
internal color sums have been performed, while in the corresponding
kinematic numerators the loop momenta are not integrated but held
fixed.  In our relatively simple one- and two-loop cases, the substitutions
(\ref{N4Replacement}) and (\ref{TwoLoopSubstitution}) hold because
they happen to be equivalent to making the substitutions prior to
switching to a color basis.   For carrying out the three-loop 
calculation of divergences, we instead directly use \eqn{DoubleCopy}.

%%%%%%%%%%%%%%%%%%%%%%%%%%%%%%%%%%%
\section{Procedure for computation}
\label{ProcedureSection}

In this section we give our procedure for constructing the
half-maximal supergravity amplitudes and then extracting the
ultraviolet divergences.

\subsection{General construction}

%%%%%%%%%%%%%% FIGURE %%%%%%%%%%%%%
\begin{figure}
\includegraphics[scale=.55]{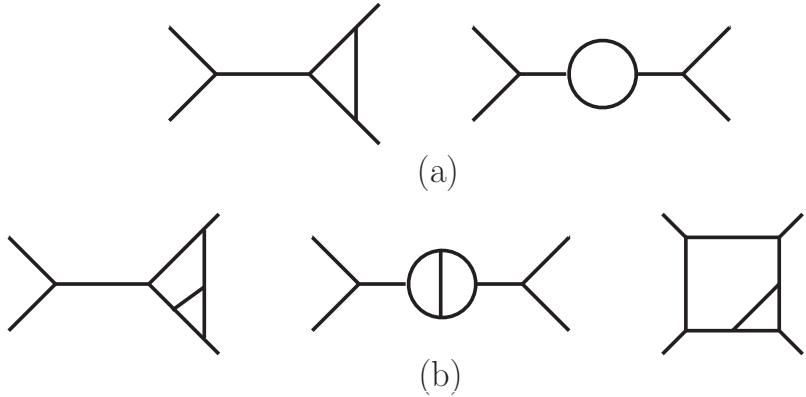}
\caption{Diagrams with triangle and bubble subgraphs at (a) one loop
  and (b) two loops.  These do not contribute to terms proportional to
  the needed color tensors in Yang-Mills and therefore do not
  contribute to the supergravity divergences.}
\label{VanishingDiagramsFigure}
\end{figure}
%%%%%%%%%%%%%%%%%%%%%%%%%%%

Using \eqns{FourPointSupergravity}{TwoLoopSugraBCJ}, we obtain one- and
two-loop half-maximal supergravity amplitudes directly from
nonsupersymmetric gauge-theory amplitudes.  Because we are interested
in cases where no integrated results exist for the amplitudes, we use
slightly modified forms where we replace the gauge-theory amplitudes
by their Feynman diagrams.  Although it may seem inefficient to
use Feynman diagrams, in our case it makes little difference because
we are interested in ultraviolet divergences in only the relatively
small number of contributions that carry the color factors of the box
diagrams at one loop and the double-box diagrams at two loops.  In
addition, we need expressions valid in general dimensions, making it
more difficult to use more sophisticated helicity methods.

As we already discussed in \sect{AmplitudesSubsection}, the gauge-theory 
divergences that feed into half-maximal supergravity
divergences are those with color factors depending on the independent color
tensor $b_1^{\oneloop}$ at one loop and the independent color tensors
$b_1^{\twoloop}$ and $b_2^{\twoloop}$ at two loops.  Any Feynman diagram that
has a triangle or bubble subgraph, as displayed in
\fig{VanishingDiagramsFigure}, will not contribute to the needed color
tensors and will therefore not contribute to the supergravity
divergences. One can also see that the antisymmetry of the kinematic
part of the vertices will cause these diagrams to cancel in the
permutation sum in \eqn{FourPointSupergravity}.

This allows us to replace the one-loop amplitudes in
\eqn{FourPointSupergravity} with box diagram contributions:
\begin{equation}
\mathcal{M}^{\oneloop}_{Q=16} =
i\Bigl(\frac{\kappa}{2}\Bigr)^4 s t A^{\rm tree}_{Q=16}(1,2,3,4)
\Big[ B^{\oneloop}(1,2,3,4) + B^{\oneloop}(1,3,4,2) + B^{\oneloop}(1,4,2,3) \Big]\,,
\hskip .5 cm 
\label{FourPointSupergravityDiagrams}
\end{equation} 
where $B^{\oneloop}(1,2,3,4)$ collects all Feynman-diagram contributions which
have the color factor of the box in \fig{BoxFigure}.  This includes
the box Feynman diagrams, whether containing scalars, ghosts or
gluons, and any terms in diagrams with four-point contact terms
carrying the box color factor.  The other contributions
$B^{\oneloop}(1,3,4,2)$ and $B^{\oneloop}(1,4,2,3)$ are similar except the
external legs are permuted.  Due to the color Jacobi relations, there
is an ambiguity in assigning terms to diagrams, but in the end it does
not matter because if a term cancels in
\eqn{FourPointSupergravityDiagrams} in one particular arrangement, it
will cancel in other arrangements as well.

At two loops the situation is similar. Expressing the gauge-theory
amplitudes in \eqn{TwoLoopSugraBCJ} in terms of Feynman diagrams, 
we find that only those diagrammatic contributions that carry the color factor
of either the planar or nonplanar double box do not cancel.   Keeping
these contributions, we have the supergravity amplitude as
\begin{eqnarray}
\mathcal{M}^{\twoloop}_{Q=16}(1,2,3,4) &=&
i\Bigl(\frac{\kappa}{2}\Bigr)^6 s t A^{\rm tree}_{Q=16}(1,2,3,4)
 \hskip -0.3 cm \sum_{x \in \{\rm P, NP\}} \Bigl[
 s (B^{x}(1,2,3,4) + B^{x}(3,4,2,1)  ) \nonumber\\
&& \hskip 0.4 cm
+ t ( B^{x}(1,4,2,3)  + B^{x}(2,3,4,1))
+ u ( B^{x}(1,3,4,2)  + B^{x}(4,2,3,1)) \Bigr]\,, \nonumber\\ 
\label{TwoLoopSugraBCJDiagrams}
\end{eqnarray}
where $B^{\rm P}(1,2,3,4)$ are the diagrammatic contributions with the
planar double-box color factor shown in \fig{DoubleBoxFigure}(a), and
$B^{\rm NP}(1,2,3,4)$ are the diagrammatic contributions containing
the nonplanar double-box color factor in \fig{DoubleBoxFigure}(b).  
The other nonvanishing contributions carry color factors
that are just relabelings of these, while all contributions 
that do not carry such color factors cancel in \eqn{TwoLoopSugraBCJDiagrams}.  As for one
loop, the assignment of the terms in each of these contributions is
not unique.

We have numerically confirmed, using helicity states in four dimensions,
that \eqn{FourPointSupergravityDiagrams} has the correct two-particle
unitarity cuts and that \eqn{TwoLoopSugraBCJDiagrams} has the correct
double two-particle cuts. As noted earlier, the one- and two-loop
double-copy formulas (\ref{FourPointSupergravityDiagrams}) and
(\ref{TwoLoopSugraBCJDiagrams}) are guaranteed to hold, as long as the
input gauge-theory amplitudes have the correct cuts.  Nevertheless,
this is a nontrivial consistency check to show that we have assembled the
contributions correctly.

At three loops the situation is somewhat more complicated.  We will
follow the construction in Ref.~\cite{N4gravThreeLoops}, where all
four-point three-loop half-maximal pure supergravity divergences were
constructed.  Here the construction is identical except that on the
nonsupersymmetric gauge-theory side of the double copy we include
scalars, giving us supergravity amplitudes including matter
multiplets.

\subsection{Dimensional reduction for matter multiplets}
\label{DimensionalReductionSubsection}

Half-maximal supergravity in $D$ dimensions with $n_\subV$ abelian
matter multiplets is conveniently generated by dimensionally reducing
pure half-maximal supergravity from ${D + n_\subV}$ dimensions to $D$
dimensions (with $D+n_{\subV} \le 10$)~\cite{deRoo,Awada,RaduMatter}.
This automatically generates half-maximal supergravity with proper
interactions between the different vector multiplets.  Indeed, in
Ref.~\cite{deRoo} the Lagrangian of $\NeqFour$ supergravity in four
dimensions with six vector multiplets is constructed via dimensional
reduction of pure ${\cal N} = 1$, $D=10$ supergravity.

This observation makes it straightforward to modify previous
computations in pure half-maximal
supergravity~\cite{N4gravThreeLoops,HalfMax5D} to now include abelian
matter multiplets.  Indeed, dimensional reduction is very natural in
the double-copy formalism. Under dimensional reduction the number of
states is unchanged; in particular, maximal super-Yang-Mills theory is just
the dimensional reduction of ${\cal N} = 1$, $D=10$ super-Yang-Mills theory.
Under dimensional reduction from $D+n_{\subV}$ dimensions to $D$
dimensions, each gluon carries $D+n_{\subV}-2$ physical states that
split into $n_{\subV}$ scalars and $D-2$ gluon states.  The tensor
product of the states of maximally supersymmetric Yang-Mills theory with a
scalar state gives a vector matter multiplet, while the tensor product
with a gluon state gives a graviton multiplet.  Therefore tensoring
dimensionally reduced nonsupersymmetric Yang-Mills theory with maximal
super-Yang-Mills theory yields half-maximal supergravity with vector matter
multiplets.

Besides the standard gauge-theory couplings, the scalars generated by
dimensional reduction in Yang-Mills theory can interact with other scalars.
To determine the appropriate scalar couplings needed for the
double-copy construction of half-maximal supergravity with matter
multiplets, we simply track the scalar interactions under
dimensional reduction.  Explicitly, under dimensional reduction we
obtain the gauge-theory Lagrangian,
%%%%%%%%%%%%%%%%%%%%%%%%%%
\begin{equation}
\mathcal{L} = \mathcal{L}_{\text{YM}} + \mathcal{L}_{\text{ghost}}
  + \mathcal{L}_{\text{scalar}}\,,
\end{equation}
%%%%%%%%%%%%%%%%%%%%%%%%%%
where the gluon, ghost, and scalar contributions are
%%%%%%%%%%%%%%%%%%%%%%%%%%
\begin{align}
\mathcal{L}_{\text{YM}}  &= 
   -\tfrac{1}{2} (\partial_\mu A_\nu^a)(\partial^\mu A^{\nu a}) 
   + \frac{ig}{\sqrt{2}} \tilde{f}^{abc} (\partial_\mu A_\nu^a)A^{\mu b}A^{\nu c} 
   + \frac{g^2}{8} \tilde{f}^{abe}\tilde{f}^{ecd} A_\mu^a A_\nu^b A^{\mu c} A^{\nu d}
        \,,  \nn \\
\mathcal{L}_{\text{ghost}} &= (\partial_\mu \bar{c}^a) (\partial^\mu c^a) 
   - \frac{ig}{\sqrt{2}} \tilde{f}^{abc} (\partial_\mu \bar{c}^a) A^{\mu b} c^c \,, 
\label{GaugeTheoryLagrangian}\\
\mathcal{L}_{\text{scalar}} &= 
    \tfrac{1}{2} (\partial_\mu \phi_{i}^a)(\partial^\mu \phi_i^a) 
  - \frac{ig}{\sqrt{2}} \tilde{f}^{abc} (\partial_\mu \phi_i^a) A^{\mu b} \phi_i^c
  - \frac{g^2}{8} \tilde{f}^{abe} \tilde{f}^{ecd}
   (2 A_\mu^a \phi_i^b A^{\mu c} \phi_i^d - \phi_i^a \phi_j^b \phi_i^c \phi_j^d )
   \,. \nn
\end{align}
%%%%%%%%%%%%%%%%%%%%%%%%%
We use Feynman gauge in $\mathcal{L}_{\text{YM}}$, which makes it
straightforward to identify the propagators in \eqn{LoopGauge}. The
scalar Lagrangian $\mathcal{L}_{\text{scalar}}$ is the result of
dimensionally reducing $\mathcal{L}_{\text{YM}}$ by separating out the
higher-dimensional components of $A$ and $\partial$ as
%%%%%%%%%%%%%%%%%%%%%%%%%%
\begin{equation}
	A_\mu^a \rightarrow (A_\mu^a, \phi_i^a) \,, \qquad  A^{\mu a} \rightarrow (A^{\mu a}, -\phi_i^a) \,, \qquad \partial_\mu \rightarrow (\partial_\mu, 0)\,.
\end{equation}
%%%%%%%%%%%%%%%%%%%%%%%%%% 
In our metric convention, we have
$\phi^{ia} = -\phi_i^a$.  Our color factors are
rescaled as in \eqn{fabcdef}.

In the bare Lagrangian (\ref{GaugeTheoryLagrangian}), we have normalized the
four-scalar interaction to carry the same coupling as the gluons.  Of
course, under renormalization the coefficient of the four-scalar
interaction is no longer locked to the gauge coupling by gauge
invariance.  Nor is the color structure locked to the one of
Yang-Mills theory.  Because the scalars and gluons of this theory have
different ultraviolet-divergence structure, the double-copy property implies
that amplitudes with external matter multiplets will also behave
differently. This has important ramifications for the
divergence structure of the double-copy supergravity theories.

As a practical matter, it is easier to not use
$\mathcal{L}_{\text{scalar}}$ explicitly but instead to incorporate
the scalars into the gluon Lagrangian $\mathcal{L}_{\text{YM}}$ taken
in $D_s=D+n_\subV$ dimensions but with all momenta restricted to the
$(D-2\eps)$-dimensional subspace (where we take $D$ to be an integer and
$\eps<0$ for the purposes of determining Lorentz dot products).  In a given
Feynman diagram, whenever the gluon propagators contract around a
loop, we take the circulating states to be in $D_s$ dimensions; in
other words, we take $\eta^\mu_{\phantom{\mu}\mu}=D_s$ assuming the
contraction is formed only from $\eta_{\mu\nu}$'s explicitly appearing
in the Feynman rules. (If a contraction is formed using also an
$\eta_{\mu\nu}$ from reducing tensor loop integrals to scalar
integrals, then the contraction instead gives $d\equiv D-2\eps$.)  In
addition, for an external scalar state, say on leg 1, we take the
polarization vector to be orthogonal to the $(D-2\eps)$-dimensional
subspace where momenta live:
\begin{equation}
\varepsilon^\phi_{1\mu} \rightarrow (0,\varepsilon_{1i})\,,
\end{equation}
so that it is annihilated whenever it contracts with a momentum vector:
\begin{equation}
\varepsilon_1 \cdot p_i = \varepsilon_1 \cdot \ell_i = 0\,.
\end{equation}
This is the crucial difference between a scalar and vector
contribution, and this difference alters the results so that color
tensors that are forbidden in the divergences of gluonic amplitudes
can appear in amplitudes with external scalars.  The only possible
nonvanishing contractions for the scalar polarization vectors are
those with the polarization vectors of other external scalars; for
example if we desire particle 2 to be another scalar, then
$\varepsilon_1\cdot\varepsilon_2$ can be nonvanishing.

\begin{figure}[tb]
\begin{center}
\includegraphics[scale=0.45]{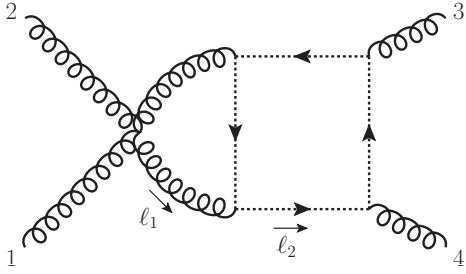}
\caption{A Feynman diagram appearing in the calculation of the 
gauge-theory amplitude $\mathcal{A}^{\twoloop}(1_\phi,2_\phi,3_g,4_g)$
and supergravity amplitude $\mathcal{M}^{\twoloop}(1_\subV,2_\subV,3_\subH,4_\subH)$.
The closed loop on the right is that of a ghost.
\label{TwoLoopContributionFigure}  }
\end{center}
\end{figure}

As a concrete example, consider the two-loop ultraviolet divergence in
five dimensions of the supergravity amplitude with two legs from a
matter multiplet and two legs from a graviton multiplet,
$\left. \mathcal{M}^{\twoloop}(1_\subV,2_\subV,3_\subH,4_\subH)
\right|_{D=5 \text{ div.}}$, where the subscripts $\rm V$ and $\rm
H$ indicate whether a leg is a state from a vector multiplet or a
graviton multiplet.  As discussed in \sect{AmplitudesSubsection}, the
super-Yang-Mills side of the double copy is incorporated by a simple
replacement of a color factor with a numerator factor.  On the
nonsupersymmetric gauge-theory side, we must compute the divergence,
$\left. \mathcal{A}^{\twoloop}(1_\phi,2_\phi,3_g,4_g)\right|_{D=5 \text{
    div.}}$, or more specifically its terms proportional to the
independent two-loop color tensors.  One contribution to this
divergence comes from the Feynman diagram shown in
\fig{TwoLoopContributionFigure} after dimensional reduction from $D_s$
dimensions.

%%%%%%%%%%%%%%%%%%%%%%%%%%
%%%%%%%%%%%%%%%%%%%%%%%%%%

This diagram involves a contact vertex and a ghost loop, and it has a
piece proportional to the color factor $c_{1234}^{\text{P}}$ shown in
\fig{DoubleBoxFigure}(a), on which we will focus in this example.
In $D_s$ dimensions, the gauge-theory integrand is given by
%%%%%%%%%%%%%%%%%%%%%%%%%%
\begin{equation}
\frac{ig^6}{8} c_{1234}^{\text{P}} 
\frac{ 
\varepsilon_{1\mu_1} \varepsilon_{2\mu_2} \varepsilon_{3\mu_3} \varepsilon_{4\mu_4}   
	(\ell_2+p_{12})^{\mu_3}
	(\ell_2-p_4)^{\mu_4}
	(\ell_2-\ell_1)_{\mu_5}
	\ell_{2\mu_6}
	(\eta^{\mu_1\mu_5}\eta^{\mu_2\mu_6} - \eta^{\mu_1\mu_2}\eta^{\mu_5\mu_6})      }
        {(\ell_1)^2   (\ell_1+p_{12})^2     (\ell_1-\ell_2)^2  
        (\ell_2)^2 (\ell_2+p_{12})^2   (\ell_2-p_4)^2 }\,,
\label{SampleIntegral}
\end{equation} 
%%%%%%%%%%%%%%%%%%%%%%%%%%
where the $p_i$ are the momenta of the external legs, $p_{12} = p_1 +
p_2$ and the $\ell_i$ are the loop momenta as indicated in
\fig{TwoLoopContributionFigure}.  The supergravity integrand is
obtained with the simple replacement, $g^6 c_{1234}^{\text{P}}
\rightarrow i(\kappa/2)^6 s^2 t A^\tree_{Q=16}$.  To obtain the
amplitude with legs $1$ and $2$ being identical scalars in gauge
theory or vector multiplet states in supergravity, we restrict the
momenta to be orthogonal to the polarization vectors $\varepsilon_1$
and $\varepsilon_2$ which live entirely in the $(D_s-D)$-dimensional
subspace.  In this way their only nonvanishing contraction is
$\varepsilon_1\cdot \varepsilon_2 = -1$ since legs $3$ and $4$ are
gluons and their polarizations live in $D$-dimensional subspace which is
not orthogonal to the momenta.  Under this restriction,
the sample in \eqn{SampleIntegral} becomes
%%%%%%%%%%%%%%%%%%%%%%%%%%
\begin{equation}
\mathcal{I}^{\,\text{sample}}_{d} = \frac{i}{8}
     \int \frac{d^d \ell_1}{(2\pi)^{d}} \frac{d^d \ell_2}{(2\pi)^d} 
	\frac{ \varepsilon_{3}\cdot(\ell_2+p_{12})  \,
	\varepsilon_{4} \cdot 	\ell_2 \,
	(\ell_2-\ell_1)\cdot \ell_{2}      }
	 {(\ell_1)^2   (\ell_1+p_{12})^2    
     (\ell_1-\ell_2)^2   (\ell_2)^2 (\ell_2+p_{12})^2   (\ell_2-p_4)^2 }\,,
\label{eq:SampleInt1}
\end{equation} 
%%%%%%%%%%%%%%%%%%%%%%%%%%
where we have integrated over $d = D - 2\eps$ with $D$ an integer and
not included the $g^6 c_{1234}^{\text{P}}$ prefactor.
The remaining task is to evaluate integrals of this type in order to extract
their ultraviolet divergences.

\subsection{Series expansion of the integrand}

Rather than evaluate integrals with their full momentum dependence,
it is advantageous to series expand the integrals to pick up 
only the desired ultraviolet divergences.  To do so we follow
the procedure of Ref.~\cite{MarcusSagnotti}.
As a first example, consider the two-loop $D=5$ case.  Odd dimensions are
a bit simpler at two loops than even dimensions because there
are never one-loop subdivergences in dimensional regularization, even
integral by integral.  Since there are no subdivergences, the $D=5$
ultraviolet divergence of the integral in \eqn{eq:SampleInt1} begins
at $\mathcal{O}(\epsilon^{-1})$ instead of
$\mathcal{O}(\epsilon^{-2})$ and is a polynomial in external momenta.  Power counting shows this polynomial to be quadratic.  We may therefore apply the dimension-counting operator
which effectively extracts powers of external momenta from the integral,
reducing its degree of divergence:
%%%%%%%%%%%%%%%%%%%%%%%%%%
\begin{equation}
\biggl(\sum_{i=1}^{4} p_{i\mu} \frac{\partial}{\partial p_{i\mu}}\biggr) 
\mathcal{I}^{\,\text{sample}}_{d=5-2\eps}  
= 2 \, \mathcal{I}^{\,\text{sample}}_{d=5-2 \eps} + \mathcal{O}(\epsilon^0) \,.
\label{eq:DimensionOperator}
\end{equation}
%%%%%%%%%%%%%%%%%%%%%%%%%% 
We use this observation to repeatedly extract powers of external
momenta from the integral, until eventually we are left with
logarithmically divergent integrals whose divergences no longer
depend on the external momenta. Explicitly, after the first
application of \eqn{eq:DimensionOperator}, we are left with
%%%%%%%%%%%%%%%%%%%%%%%%%%
\begin{eqnarray}
&& \null \hskip -.5 cm 
2 \, \mathcal{I}^{\,\text{sample}}_{d=5-2\eps} +\mathcal{O}(\epsilon^0) \nonumber \\
& & \hskip .6 cm \null 
 = \frac{i}{8} \int \frac{d^{5-2\epsilon} \ell_1}{(2\pi)^{5-2\epsilon}}
           \frac{d^{5-2\epsilon} \ell_2}{(2\pi)^{5-2\epsilon}} 
	\frac{ 	\varepsilon_{4} \cdot 	\ell_2 \,
	 (\ell_2-\ell_1)\cdot \ell_{2}      }
	 {(\ell_1)^2   (\ell_1+p_{12})^2     (\ell_1-\ell_2)^2   (\ell_2)^2 (\ell_2+p_{12})^2   (\ell_2-p_4)^2 } \nonumber\\
& &\null \hskip 1.1 cm  \times 
 \biggl\{
\varepsilon_3\cdot p_{12} - 2\varepsilon_3\cdot (\ell_2+p_{12}) \biggl(
	\frac{p_{12}\cdot \ell_1+s_{12}}{(\ell_1+p_{12})^2} +
	\frac{p_{12}\cdot\ell_2+s_{12}}{(\ell_2+p_{12})^2} -
	\frac{p_4\cdot \ell_2}{(\ell_2-p_4)^2}\biggr)\biggr\} \,.
    \hskip 1 cm 
\end{eqnarray}
%%%%%%%%%%%%%%%%%%%%%%%%%% 
We see that the quadratically divergent
term in \eqn{eq:SampleInt1} has been eliminated in favor of linearly
and logarithmically divergent integrals, along with some ultraviolet-finite
terms which we ignore. One more application of
\eqn{eq:DimensionOperator} yields, after some rearrangement and
dropping of finite pieces, the result,
%%%%%%%%%%%%%%%%%%%%%%%%%%
\begin{eqnarray}
&& \null \hskip -.5 cm 
\mathcal{I}^{\,\text{sample}}_{d=5-2\epsilon} + \mathcal{O}(\epsilon^0) \nn \\
&& \hskip 1cm \null
   =  \frac{i}{8} \int \frac{d^{5-2\epsilon} \ell_1}{(2\pi)^{5-2\epsilon}}
       \frac{d^{5-2\epsilon} \ell_2}{(2\pi)^{5-2\epsilon}} 
	\frac{ \varepsilon_{4} \cdot 	\ell_2 \,
	(\ell_2-\ell_1)\cdot \ell_{2}      }
	 {(\ell_1)^2   (\ell_1+p_{12})^2     
         (\ell_1-\ell_2)^2 (\ell_2)^2 (\ell_2+p_{12})^2 
          (\ell_2-p_4)^2 } \nonumber \\
&& \hskip 1.5 cm \null
  \times \biggl\{
	-\varepsilon_3\cdot p_{12}\, X
	-\varepsilon_3\cdot\ell_2 \biggl(\frac{s_{12}}
         {(\ell_1+p_{12})^2} + \frac{s_{12}}{(\ell_2+p_{12})^2} \biggr) \nn \\
&& \hskip 2.cm  \null 
		+\frac{1}{2}\varepsilon_3\cdot\ell_2 \biggl(
		\frac{(2p_{12}\cdot\ell_1)^2}{(\ell_1+p_{12})^4} 
		+\frac{(2p_{12}\cdot\ell_2)^2}{(\ell_2+p_{12})^4} 
		+\frac{(2p_{4}\cdot\ell_2)^2}{(\ell_2-p_{4})^4}
				+X^2\biggr) \biggr\} \,, \hskip 1 cm
\end{eqnarray}
%%%%%%%%%%%%%%%%%%%%%%%%%%
where we have defined the quantity,
%%%%%%%%%%%%%%%%%%%%%%%%%%
\begin{equation}
X = \frac{2p_{12}\cdot\ell_1}{(\ell_1+p_{12})^2} + \frac{2p_{12}\cdot \ell_2}{(\ell_2+p_{12}^2)} - \frac{2p_4\cdot\ell_2}{(\ell_2-p_4)^2}\,.
\end{equation}
%%%%%%%%%%%%%%%%%%%%%%%%%% 
At this point, the sample integral is purely
logarithmically divergent, and the polynomial dependence of its
divergence is manifest. We can now freely alter the
dependence in the propagators on external momenta without worrying
about affecting the divergence. In particular, we can take $p_i
\rightarrow 0$ in the propagators. As a result, we see that what we
have done to the original integral is equivalent to making the
propagator replacements,
%%%%%%%%%%%%%%%%%%%%%%%%%%
\begin{equation}
	\frac{1}{(\ell-p)^2} \rightarrow \frac{1}{\ell^2} \sum_{n=1}^{\infty}
\biggr(\frac{2\ell\cdot p-p^2}{\ell^2}\biggr)^{\! n}\,,
\end{equation}
%%%%%%%%%%%%%%%%%%%%%%%%%%
and retaining only the logarithmically divergent terms. 

Taking $p_i\rightarrow 0$ in the propagators makes the integrals much
simpler, but then they no longer have a scale and technically vanish in
dimensional regularization. To correct this, we must re-introduce a
scale. A computationally convenient choice is to give all of the
propagators a uniform mass $m$. This makes the integral well defined
and more tractable:
%%%%%%%%%%%%%%%%%%%%%%%%%%
\begin{eqnarray}
 \mathcal{I}^{\,\text{sample}}_{d=5-2\epsilon} + \mathcal{O}(\epsilon^0) 
&=&  \frac{i}{8} \int \frac{d^{5-2\epsilon} \ell_1}{(2\pi)^{5-2\epsilon}} \frac{d^{5-2\epsilon} \ell_2}{(2\pi)^{5-2\epsilon}} 
	\frac{ 	\varepsilon_{4} \cdot 	\ell_2 \,
	(\ell_2-\ell_1)\cdot \ell_{2}      }
	{(\ell_1^2-m^2)^2  ((\ell_1-\ell_2)^2-m^2) (\ell_2^2-m^2)^3 } \nonumber \\
&&\hskip 0.7cm  \null  \times 
      \biggl\{ -\varepsilon_3\cdot p_{12}\, \tilde{X}
	       -\varepsilon_3\cdot\ell_2 \left(\frac{s_{12}}{\ell_1^2-m^2} + \frac{s_{12}}{\ell_2^2-m^2}  \right) \label{eq:SampleInt2} \\
&&\hskip 1.5 cm  \null
	+\frac{1}{2}\varepsilon_3\cdot\ell_2 \left(
	\frac{(2p_{12}\cdot\ell_1)^2}{(\ell_1^2-m^2)^2} 
	+\frac{(2p_{12}\cdot\ell_2)^2}{(\ell_2^2-m^2)^2} 
	+\frac{(2p_{4}\cdot\ell_2)^2}{(\ell_2^2-m^2)^2}
	+\tilde{X}^2\right) \biggr\} \,, \nonumber \hskip .5cm 
\end{eqnarray}
%%%%%%%%%%%%%%%%%%%%%%%%%%
with
%%%%%%%%%%%%%%%%%%%%%%%%%%
\begin{equation}
\tilde{X} = \frac{2p_{12}\cdot\ell_1}{\ell_1^2-m^2} + \frac{2p_{12}\cdot \ell_2}{\ell_2^2-m^2} - \frac{2p_4\cdot\ell_2}{\ell_2^2-m^2}\,.
\end{equation}
%%%%%%%%%%%%%%%%%%%%%%%%%%

\subsection{Tensor reduction}

The next step in the analysis of
$\mathcal{I}^{\,\text{sample}}_{d=5-2\epsilon}$ is to simplify the
tensor numerators.  This can be handled straightforwardly using
Lorentz invariance, as recently discussed in, for example,
Ref.~\cite{Neq44np}.  Consider the terms in \eqn{eq:SampleInt2}
proportional to
%%%%%%%%%%%%%%%%%%%%%%%%%%
\begin{equation}
\mathcal{I}_{\text{tensor}}^{\mu_1\mu_2\mu_3\mu_4} = 
\int \frac{d^{5-2\epsilon} \ell_1}{(2\pi)^{5-2\epsilon}} \frac{d^{5-2\epsilon} \ell_2}{(2\pi)^{5-2\epsilon}}
\frac{\ell_2^{\mu_1} \ell_2^{\mu_2} \ell_1^{\mu_3}\ell_1^{\mu_4}\, (\ell_2-\ell_1)\cdot \ell_2}
	 {(\ell_1^2-m^2)^4  ((\ell_1-\ell_2)^2-m^2)   (\ell_2^2-m^2)^3 }\,.
\end{equation}
%%%%%%%%%%%%%%%%%%%%%%%%%%
This is a rank-4 tensor integral, but because no dependence on 
the external momenta remains, it must evaluate to a linear combination 
of products of metric tensors, as nothing else is available:
%%%%%%%%%%%%%%%%%%%%%%%%%%
\begin{equation}
\mathcal{I}_{\text{tensor}}^{\mu_1\mu_2\mu_3\mu_4} =  
	\alpha_1 \eta^{\mu_1\mu_2}\eta^{\mu_3\mu_4}
	+\alpha_2 \eta^{\mu_1\mu_3}\eta^{\mu_2\mu_4}
	+\alpha_3 \eta^{\mu_1\mu_4}\eta^{\mu_2\mu_3}\,.
\label{eq:TensorReduction1}
\end{equation}
%%%%%%%%%%%%%%%%%%%%%%%%%%
The particular integrand of $\mathcal{I}_{\text{tensor}}^{\mu_1\mu_2\mu_3\mu_4}$ enforces a symmetry between $\mu_1\leftrightarrow \mu_2$ and $\mu_3\leftrightarrow \mu_4$, so that $\alpha_2=\alpha_3$, but we will ignore such optimizations here. Instead, we contract the indices of \eqn{eq:TensorReduction1} in all possible ways to obtain the following system of three equations:
%%%%%%%%%%%%%%%%%%%%%%%%%%
\begin{align}
\eta_{\mu_1\mu_2}\eta_{\mu_3\mu_4}\mathcal{I}_{\text{tensor}}^{\mu_1\mu_2\mu_3\mu_4} &=  
	\alpha_1 (5-2\epsilon)^2 + \alpha_2  (5-2\epsilon) + \alpha_3  (5-2\epsilon)\,, \nonumber \\
	\eta_{\mu_1\mu_3}\eta_{\mu_2\mu_4}\mathcal{I}_{\text{tensor}}^{\mu_1\mu_2\mu_3\mu_4} &=  
	\alpha_1  (5-2\epsilon) + \alpha_2  (5-2\epsilon)^2 + \alpha_3  (5-2\epsilon)\,, \nonumber \\
	\eta_{\mu_1\mu_4}\eta_{\mu_2\mu_3}\mathcal{I}_{\text{tensor}}^{\mu_1\mu_2\mu_3\mu_4} &=  
	\alpha_1  (5-2\epsilon) + \alpha_2  (5-2\epsilon) + \alpha_3  (5-2\epsilon)^2\,.
\end{align}
%%%%%%%%%%%%%%%%%%%%%%%%%%
The left-hand sides of these equations are
scalar, single-scale vacuum integrals, which are amenable to direct
integration. So to evaluate
$\mathcal{I}_{\text{tensor}}^{\mu_1\mu_2\mu_3\mu_4}$, we just need to
invert these equations and solve for $\alpha_i$.  The same 
idea works as well for higher-rank tensors, allowing us to 
reduce all the tensor integrals to scalar integrals.

After performing the tensor reduction, it is useful to cancel 
as many propagators as possible using numerator replacements,
%%%%%%%%%%%%%%%%%%%%%%%%%%
\begin{align}
\ell_1^2 &\rightarrow (\ell_1^2-m^2)+m^2\,,\nonumber \\
 \ell_2^2 &\rightarrow (\ell_2^2-m^2)+m^2\,,\nonumber\\
\ell_1\cdot \ell_2 &\rightarrow -\frac{1}{2}\Bigl(((\ell_1-\ell_2)^2-m^2) - (\ell_1^2-m^2) - (\ell_2^2-m^2) - m^2\Bigr)\,.
\end{align}
%%%%%%%%%%%%%%%%%%%%%%%%%%
In this way the divergence of the integral
$\mathcal{I}^{\,\text{sample}}_{d=5-2\epsilon}$ becomes a linear
combination of scalar, single-scale vacuum integrals of the form,
%%%%%%%%%%%%%%%%%%%%%%%%%%
\begin{equation}
\int \frac{d^{5-2\epsilon} \ell_1}{(2\pi)^{5-2\epsilon}} \frac{d^{5-2\epsilon} \ell_2}{(2\pi)^{5-2\epsilon}} \frac{1}{(\ell_1^2-m^2)^{a_1} ((\ell_1-\ell_2)^2-m^2)^{a_2} (\ell_2^2-m^2)^{a_3}}\,,
\label{eq:scalarInt1}
\end{equation}
%%%%%%%%%%%%%%%%%%%%%%%%%%
with $a_i$ integers. 

\subsection{Scalar integral evaluation}

%%%%%%%%%%%% TABLE %%%%%%%%%%%%%%%%%%%%%%%%%%
\begin{table*}[ht]\caption{
The basis integrals in $d=4-2\epsilon,\,5-2\epsilon,\,6-2\epsilon$
required for the two-loop Yang-Mills and gravity divergence
computations, valid through ${\cal O} (1/\epsilon)$.  A factor of
$1/(4\pi)^d$ has been dropped from the results in the table.
\label{IntegralTable} }
\vskip .4 cm
\begin{center}
\resizebox{16cm}{!}{\begin{tabular}{||c||c|c|c||}
\hline
Integral & $d=4-2\epsilon$ & $d=5-2\epsilon$ & $d=6-2\epsilon$ \\
\hline
\hline
\multirow{2}{*}{$\mathcal{I}_1$} & \multirow{2}{*}{\raisebox{1pt}{$-$}{\large{$\frac{1}{\epsilon^2}$}}$(m^2)^{2-2\epsilon}e^{2(1-\gamma_{\rm{E}})\epsilon}+\mathcal{O}(\epsilon^0)$} & \multirow{2}{*}{$\mathcal{O}(\epsilon^0)$} & \multirow{2}{*}{\raisebox{1pt}{$-$}{\large{$\frac{1}{4\epsilon^2}$}}$(m^2)^{4-2\epsilon}e^{(3-2\gamma_{\rm{E}})\epsilon}+\mathcal{O}(\epsilon^0)$} \\
& & & \\
\hline
\multirow{2}{*}{$\mathcal{I}_2$} & \multirow{2}{*}{\raisebox{.8pt}{$-$}{\large{$\frac{3}{2\epsilon^2}$}}$(m^2)^{1-2\epsilon}e^{(3-2\gamma_{\rm{E}})\epsilon}+\mathcal{O}(\epsilon^0)$} & \multirow{2}{*}{{\large{$\frac{\pi}{2\epsilon}$}}$(m^2)^2+\mathcal{O}(\epsilon^0)$} & \multirow{2}{*}{\raisebox{.8pt}{$-$}{\large{$\frac{5}{8\epsilon^2}$}}$(m^2)^{3-2\epsilon}e^{(113/30-2\gamma_{\rm{E}})\epsilon}+\mathcal{O}(\epsilon^0)$} \\
& & & \\
\hline
\end{tabular}}
\end{center}
\end{table*}
%%%%%%%%%%%%%%%%%%%%%%%%%%%%%%%%%%%%%%%%%%%%%%%%%

After reducing $\mathcal{I}^{\,\text{sample}}_{d=5-2\epsilon}$ to scalar integrals of the form~\eqref{eq:scalarInt1}, we must evaluate the integrals.  We first reduce them to a basis using integration by parts as implemented in \verb|FIRE|~\cite{Fire}.  In all dimensions considered here, our basis consists of two scalar vacuum integrals:
\begin{align}
&\mathcal{I}_1=\int \frac{d^{d} \ell_1}{(2\pi)^{d}} 
\frac{d^{d} \ell_2}{(2\pi)^{d}} \frac{1}{(\ell_1^2-m^2)(\ell_2^2-m^2)}\,,
 \notag \\
&\mathcal{I}_2=\int \frac{d^{d} \ell_1}{(2\pi)^{d}}
 \frac{d^{d} \ell_2}{(2\pi)^{d}} 
\frac{1}{(\ell_1^2-m^2)((\ell_1-\ell_2)^2-m^2)(\ell_2^2-m^2)}\,.
\end{align}
The first integral is simply a product of two easily evaluated
one-loop integrals.  We evaluate the second integral using the code
\verb|MB|~\cite{MB} that implements Mellin-Barnes
integration~\cite{MellinBarnes}.  The results are collected in
Table~\ref{IntegralTable}, where an overall prefactor of $1/(4\pi)^d$
has been removed for simplicity.  For the cases considered here, we
need the basis integrals through order $1/\epsilon$.  Using these
results for the scalar integrals completes the evaluation of
$\mathcal{I}^{\,\text{sample}}_{d=5-2\epsilon}$ and other similar
two-loop integrals prior to the subtraction of subdivergences.

\subsection{Subdivergences}

The example $\mathcal{I}^{\,\text{sample}}_{d=5-2\epsilon}$ is special
in that it has no subdivergences.  More generally, subdivergences
occur and can greatly complicate the analysis.  To deal with this, we
follow the basic approach of Ref.~\cite{MarcusSagnotti}.  If we alter the 
previous example to be in six dimensions instead of five dimensions, then 
we see by power counting that 
$\mathcal{I}^{\,\text{sample}}_{d=6-2\epsilon}$ in \eqn{eq:SampleInt1} has one-loop
subdivergences in both the $\ell_1$ and $\ell_2$ integrals. There is
also a third subloop that could in principle have a divergence --- the
loop parametrized by $\ell_1+\ell_2$ --- but it turns out to
be finite in $\mathcal{I}^{\,\text{sample}}_d$ for $d<8$.

The presence of subdivergences means that 
$\mathcal{I}^{\,\text{sample}}_{d=6-2\epsilon}$ begins at 
$\mathcal{O}(\epsilon^{-2})$, and \eqn{eq:DimensionOperator} will 
need to be modified.  One possible way to modify it  is that  we need to
keep factors of $\eps$ that can strike a $1/\eps^2$:
%%%%%%%%%%%%%%%%%%%%%%%%%%
\begin{equation}
\biggl(\sum_{i=1}^{4} p_{i\mu} \frac{\partial}{\partial p_{i\mu}}\biggr)
 \mathcal{I}^{\,\text{sample}}_{d=6-2\epsilon}  
= (4-4\epsilon) \, \mathcal{I}^{\,\text{sample}}_{d=6-2\epsilon} \,,
\label{eq:DimensionOperator2}
\end{equation}
%%%%%%%%%%%%%%%%%%%%%%%%%% 
which holds to all orders in $\epsilon$. However,
we cannot simply disregard terms that are naively finite by overall power
counting because they may still contain subdivergences that would
contribute at $\mathcal{O}(\epsilon^{-1})$. In addition, after 4
powers of external momenta have been extracted, leaving only
logarithmically divergent integrals and pure-subdivergence integrals,
we cannot set $p_i\rightarrow 0$ or add masses in the propagators
without affecting the $\mathcal{O}(\epsilon^{-1})$ term.  In fact,
the results would depend on the details used to regulate 
infrared singularities generated by the momentum expansion.

For these reasons, we instead work with subtracted
divergences, which we denote as $S[\mathcal{I}]$. A subtracted
divergence of an integral is the integral's divergence in dimensional
regularization with all of its subdivergences subtracted off:
%%%%%%%%%%%%%%%%%%%%%%%%%%
\begin{align}
S\biggl[\int \prod_{i=1}^{L} \frac{d^d \ell_i}{(2\pi)^d} 
I(\ell_1,\ldots,\ell_L)\bigg] 
&= \text{Div}\biggl[\int \prod_{i=1}^{L} \frac{d^d \ell_i}{(2\pi)^d} 
I(\ell_1,\ldots,\ell_L)\biggr] \label{eq:Subtractions} \\
&\hskip -1cm  - \sum_{l=1}^{L-1} \hskip -.1 cm 
       \sum_{l-\text{loop} \atop \text{subintegrals}} \hskip -.3 cm 
\text{Div}\biggl[\int \prod_{i=l+1}^{L} \frac{d^d \tilde{\ell}_i}{(2\pi)^d}  
 \, S\biggl[ \int \prod_{j=1}^l \frac{d^d\tilde{\ell}_j}{(2\pi)^d} 
  I(\tilde{\ell}_1,\ldots,\tilde{\ell}_L) \biggr]\biggr]\,. \nn
\end{align}
%%%%%%%%%%%%%%%%%%%%%%%%%% 
Here, $\text{Div}$ indicates the divergent part
of the integral (\emph{i.e.} its value through $\mathcal{O}(\epsilon^{-1})$),
and $\tilde{\ell}_i$ is a reparametrization of the integral such that
a particular $l$-loop subintegral is parametrized by $\tilde{\ell}_1$
through $\tilde{\ell}_l$. This definition can be thought of as adding
counterterm diagrams integral by integral to remove their subdivergences. 
It has the nice property
that $S[\mathcal{I}]$ is a polynomial in external momenta, so we can
extract all of the external dependence of the sample integral, in this case of quartic order, with
%%%%%%%%%%%%%%%%%%%%%%%%%%
\begin{align}
\biggl(\sum_{i=1}^{4} p_{i\mu} \frac{\partial}{\partial p_{i\mu}}\biggr)
   S[\mathcal{I}^{\,\text{sample}}_{d=6-2\epsilon} ]
		&= 4 \, S[\mathcal{I}^{\,\text{sample}}_{d=6-2\epsilon} ]\,,
\label{eq:DimensionOperator3}
\end{align}
and then freely set $p_i\rightarrow 0$ and introduce masses into the
propagators. In contrast to \eqn{eq:DimensionOperator2}, we have a $4$ instead
of a $(4- 4 \eps)$ on the right-hand side because the subtractions remove
the source of the additional terms. 
 In the case that $S[\ldots]$ is a subintegral (as in the
second line of \eqn{eq:Subtractions}), the remaining loop momenta and
any introduced masses should also be treated as external variables, so
that the dimension-counting operator is instead
%%%%%%%%%%%%%%%%%%%%%%%%%%
\begin{equation}
	\sum_{i=1}^4 p_{i\mu} \frac{\partial}{\partial p_{i\mu}} + \sum_{i=l+1}^L \tilde{\ell}_{i\mu} \frac{\partial}{\partial \tilde{\ell}_{i\mu}} + 2 m^2 \frac{\partial}{\partial m^2}\,.
\end{equation}
%%%%%%%%%%%%%%%%%%%%%%%%%%

After the subtractions are taken into account, the tensor integrals
can be simplified as before to obtain a final answer consisting of a
linear combination of scalar single-scale vacuum integrals.

%%%%%%%%%%%%%%%%%%%%%%%%%%%%%%%%%%%%%%%%%%%%%%%
\section{One-loop divergences}
\label{OneLoopDivergenceSection}

As a warm up before turning to two and three loops, we determine the one-loop 
four-point divergences of half-maximal supergravity
with $n_{\subV}$ matter multiplets in $D=4,6,8$.  (We do not
consider $D=5,7$ because there are no divergences in dimensional
regularization in odd dimensions for odd loop orders.)  We
confirm the appearance of divergences in four-matter amplitudes found
long ago by Fischler~\cite{Fischler} and by Fradkin and
Tseytlin~\cite{Arkady}.  We also illustrate the connection of
the supergravity divergences to those of four-scalar amplitudes in
nonsupersymmetric gauge theory, as noted in Ref.~\cite{HalfMax5D}.

%%%%%%%%%%%%%%%%%%%%
%FIGURE
%
\begin{figure}
\begin{center}
\includegraphics[scale=0.6]{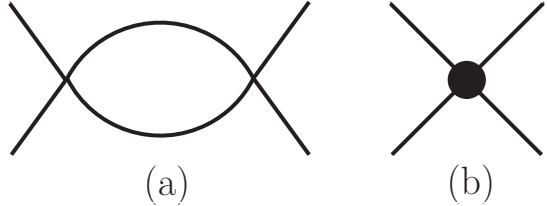}
\caption{The divergences containing a one-loop color tensor and a factor $D_s$ all arise
  from diagrams of the form (a).  The total divergence
  subtraction is shown in (b), where the large dot signifies a local
  subtraction.}
\label{DsTermsFigure}
\end{center}
\end{figure}
%%%%%%%%%%%%%%%%%%%%%%

We start by presenting the divergences in Yang-Mills theory coupled to
scalars that are proportional to the one-loop color tensor
$b_1^{\oneloop}$.  The half-maximal supergravity divergences are then
obtained by replacing the color factor with the $\mathcal{N}=4$
super-Yang-Mills BCJ numerator, as given in \eqn{N4Replacement}.  We
collect the counterterms corresponding to the supergravity divergences
with external gravitons and matter vectors in the appendix.  In $D=4$
the divergences and counterterms all carry an $SO(6) \times
SO(n_\subV)$ symmetry.  The $SO(6)$ is just the symmetry of the vectors
of the graviton multiplet and is inherited from the $R$-symmetry of
the scalars of $\NeqFour$ super-Yang-Mills theory.  The $SO(n_\subV)$
symmetry is a reflection of the fact that all matter multiplets 
are equivalent.

\subsection{Four dimensions}

As discussed in Ref.~\cite{HalfMax5D}, in $D=4$ the renormalizability
of gauge theory ensures that four-point divergences involving external
gluons must be proportional to tree-level amplitudes.  This means that
divergences proportional to the one-loop color tensor vanish:
\begin{eqnarray}
\mathcal{A}^\oneloop(1_g, 2_g, 3_g, 4_g) \bigr|_{D=4\, \mathrm{div.}} = 0 + \cdots\,,
 \nn \\ 
\mathcal{A}^\oneloop(1_g, 2_g, 3_{\phi}, 4_{\phi}) \bigr|_{D=4\,
  \mathrm{div.}} = 0 + \cdots \,,
\end{eqnarray}
where the label $g$ or $\phi$ indicates that an external leg is a
gluon or scalar, respectively, and as before, ``$+\cdots$'' signifies that we dropped divergences
proportional to the tree color tensor.

On the other hand, renormalizability
does not protect divergences in the four-scalar amplitude because
operators of the form $b_1^{\oneloop, {abcd}} \phi^a \phi^b \phi^c \phi^d$
are perfectly valid counterterms.  Carrying out the computation,
we find that for four identical external scalar states the divergence is
\begin{equation}
\mathcal{A}^{\oneloop}(1_{\phi},2_{\phi},3_{\phi},4_{\phi})
\bigr|_{D=4\,\mathrm{div.}}
=\frac{i}{\epsilon} \frac{1}{(4\pi)^2} g^4 b_1^{\oneloop}\, \frac{3(D_s-2)}{2}
+ \cdots\,,
\label{phi4OneLoop}
\end{equation}
where $D_s - 4$ is the number of distinct real scalars that
circulate in the loop.  In this case, for consistency we should take
the state-counting parameter $D_s \ge 5$ so that we have at least
one scalar state.  Taking the state-counting parameter to be an
integer which leaves the number of gluon states (for each color) at
their four-dimensional values is equivalent to using the four-dimensional
helicity scheme~\cite{FDH}.

For a pair of distinct external scalars, we find the divergence,
\begin{equation}
\mathcal{A}^{\oneloop}(1_{\phi_1},2_{\phi_1},3_{\phi_2},4_{\phi_2})
\bigr|_{D=4\,\mathrm{div.}}
=\frac{i}{\epsilon}\frac{1}{(4\pi)^2} g^4 b_1^{\oneloop} \, \frac{D_s-2}{2} + \cdots\,.
\label{phi4OneLoopDistinct}
\end{equation}
As expected, the case with identical scalars follows from the one with
distinct scalars by summing over the 3 distinct permutations
corresponding to the distinct ways of connecting the external scalar
legs.  The number of scalar states circulating in the loop is again
given by $D_s - 4$, so for consistency we should take $D_s \ge 6$ to
have two scalars.  In both \eqns{phi4OneLoop}{phi4OneLoopDistinct},
the terms containing $D_s$, and therefore those due to scalar states
in the loop, arise from contact diagrams of the form displayed in
\fig{DsTermsFigure}(a).  We will find this useful in
\sect{TwoLoopDivergenceSection} for understanding the structure of the
two-loop divergences in $D=4$.

Using the double-copy replacement (\ref{N4Replacement}) for the color
factor in terms of the $\NeqFour$ super-Yang-Mills BCJ numerator, we obtain the
corresponding divergences in $\NeqFour$ supergravity with $n_{\subV} = D_s -
4$ matter multiplets:
\begin{eqnarray}
&&\mathcal{M}^{\oneloop}(1_{\subH},2_{\subH},3_{\subH},4_{\subH})\bigr|_{D=4\,\mathrm{div.}} = 0\,, \nonumber \\
&&\mathcal{M}^{\oneloop}(1_{\subH},2_{\subH},3_{\subV},4_{\subV})\bigr|_{D=4\,\mathrm{div.}} = 0\,, \nonumber \\
  &&\mathcal{M}^{\oneloop}(1_{\subV},2_{\subV},3_{\subV},4_{\subV})\bigr|_{D=4\,\mathrm{div.}}
  =-\frac{1}{\epsilon}\frac{1}{(4\pi)^2}\Bigl(\frac{\kappa}{2}\Bigr)^4
  stA_{Q=16}^{\tree}\frac{3(D_s-2)}{2}\,, \nonumber \\
 && \mathcal{M}^{\oneloop}(1_{\subV_1},2_{\subV_1},3_{\subV_2},4_{\subV_2})\bigr|_{D=4\,\mathrm{div.}}
=-\frac{1}{\epsilon}\frac{1}{(4\pi)^2}\Bigl(\frac{\kappa}{2}\Bigr)^4
 s t A_{Q=16}^{\tree}\frac{D_s-2}{2}\,,
\label{eq:DEq4Grav}
\end{eqnarray}
where, as noted in \sect{DimensionalReductionSubsection}, the label
$\rm H$ indicates that a leg is a state of the graviton multiplet
while a subscript $\rm V$ indicates that the leg is a state of a
vector multiplet.  The cases with subscripts $\rm V_1$ and $\rm V_2$
indicate that the legs belong to distinct vector multiplets.  Cases
with an odd number of external matter multiplet legs vanish
trivially. The total number of matter vector multiplets is given by
$n_\subV = D_s -4$, and the supersymmetric prefactor
$A_{Q=16}^{\tree}$ automatically incorporates all valid external
states in both the vector and graviton multiplets. It is interesting
to note that the contribution to the divergence from the matter
multiplet in the loop is proportional to that of the graviton
multiplet, and that the result diverges for any number of vector
multiplets.  For consistency, we must have $n_{\subV} \ge 1$ for the
cases with all matter belonging to the same matter multiplet and
$n_{\subV} \ge 2$ for the case where the two pairs of external states
belong to different matter multiplets.

\subsection{Six dimensions}

As already discussed in Ref.~\cite{HalfMax5D}, the only available $F^3$
Yang-Mills counterterm for external gluons generates amplitudes with color tensors proportional
to the tree-level color tensors, a fact that is unaltered with the addition of
scalars to the theory.  Thus we immediately have that the part of the
divergence proportional to the one-loop color tensor vanishes:
\begin{equation}
{\cal A}^{\oneloop}(1_g, 2_g, 3_g, 4_g) = 0 + \cdots\,.
\label{A4DivOneloopD6}
\end{equation}

For four identical external scalars, the Yang-Mills counterterm
involving the one-loop color tensor of the form $D^2\phi^4$ vanishes
because by crossing symmetry, it needs to be proportional to
$s+t+u=0$. One might worry about an interference of the crossing
properties of the color and the kinematics, but the independent
one-loop color tensor can be put into a fully crossing-symmetric form
plus terms proportional to tree color factors:
\begin{equation}
b_{1}^{\oneloop} = \frac{1}{4!}
  \sum_\sigma c^{\oneloop}_{\sigma(1), \sigma(2), \sigma(3), \sigma(4)}  + \cdots\,,
\label{OneLoopColorSymm}
\end{equation}
where $c^{\oneloop}_{ijkl}$ is a one-loop box color factor and $\sigma$
runs over all 4! permutations of the external legs.  Therefore for
trivial symmetry reasons there is no divergence in terms containing
the one-loop color tensor when all four external scalars are
identical:
\begin{equation}
\mathcal{A}^{\oneloop}(1_{\phi},2_{\phi},3_{\phi},4_{\phi})
\bigr|_{D=6\,\mathrm{div.}} = 0 + \cdots \,.
\label{phi4DivOneloopD6}
\end{equation}
For the case of two pairs of non-identical scalars, the amplitude no
longer has the full crossing symmetry and hence the divergence no
longer vanishes from simple symmetry considerations.  Instead we find
\begin{equation}
\mathcal{A}^{\oneloop}(1_{\phi_1},2_{\phi_1},3_{\phi_2},4_{\phi_2})
   \bigr|_{D=6\,\mathrm{div.}}
=-\frac{i}{\epsilon}\frac{1}{(4\pi)^3}g^4b_1^{\oneloop}\frac{26-D_s}{12}s + \cdots \,.
\label{phi1phi1phi2phi2DivOneloopD6}
\end{equation}
Finally, the two-scalar two-gluon divergence proportional to the one-loop color
tensor is
\begin{equation}
\mathcal{A}^{\oneloop}(1_g,2_g,3_{\phi},4_{\phi})\bigr|_{D=6\,\mathrm{div.}}
=\frac{i}{\epsilon}\frac{1}{(4\pi)^3}g^4\, b_1^{\oneloop}\, \frac{26-D_s}{24}
(\varepsilon_1\cdot\varepsilon_2\,s
        -2k_1\cdot\varepsilon_2\,k_2\cdot\varepsilon_1) + \cdots \,.
\label{AAphiphiDivOneloopD6}
\end{equation}

Substituting the color factor with the kinematic numerator
(\ref{N4Replacement}) in Eqs.~(\ref{A4DivOneloopD6}),
(\ref{phi4DivOneloopD6}), (\ref{phi1phi1phi2phi2DivOneloopD6}) and
(\ref{AAphiphiDivOneloopD6}) immediately gives us the half-maximal
supergravity divergences for cases including external states from
vector multiplets: 
\begin{eqnarray}
&&\mathcal{M}^{\oneloop}(1_{\subH},2_{\subH},3_{\subH},4_{\subH})\bigr|_{D=6\,\mathrm{div.}}= 0\,,
 \nonumber \\
&&\mathcal{M}^{\oneloop}(1_{\subV},2_{\subV},3_{\subV},4_{\subV})\bigr|_{D=6\,\mathrm{div.}}= 0\,,
 \nonumber \\
 &&\mathcal{M}^{\oneloop}(1_{\subV_1},2_{\subV_1},3_{\subV_2},4_{\subV_2})\bigr|_{D=6\,\mathrm{div.}}
=\frac{1}{\epsilon}\frac{1}{(4\pi)^3}\Bigl(\frac{\kappa}{2}\Bigr)^4
  s t A_{Q =16}^{\tree}\frac{26-D_s}{12}s\,,  \label{OneLoopSugraDivD6}\\
&&\mathcal{M}^{\oneloop}(1_{\subH},2_{\subH},3_{\subV},4_{\subV})\bigr|_{D=6\,\mathrm{div.}}
= -\frac{1}{\epsilon}\frac{1}{(4\pi)^3}\Bigl(\frac{\kappa}{2}\Bigr)^4
stA_{Q=16}^{\tree}\frac{26-D_s}{24} \left(\varepsilon_1\cdot\varepsilon_2\,s
  -2k_1\cdot\varepsilon_2\,k_2\cdot\varepsilon_1\right)\,. \nonumber
\end{eqnarray}

%%%%%%%%%%%%%%%%%%%%%%%%%%%
\subsection{Eight dimensions}

In eight dimensions at one loop, nonsupersymmetric Yang-Mills theory has an
$F^4$ divergence containing a one-loop color tensor.  Therefore the
corresponding half-maximal supergravity diverges at one
loop~\cite{HalfMax5D}.  The explicit value of the divergences for four
external graviton multiplets is given in Eq.~(3.19) of
Ref.~\cite{HalfMax5D}, with the number of vector supermultiplets given by
$n_\subV = D_s - 8$; the pure supergravity divergence was first computed in
Ref.~\cite{Dunbar}. 
 Yang-Mills operators generating divergences for
external scalars in $D=8$, specifically $D^2\phi^2F^2$ and
$D^4\phi^4$, can also be contracted with one-loop color tensors.  Thus
it is no surprise that cases with external matter multiplets also
diverge in half-maximal supergravity.

For four identical scalars in Yang-Mills theory, the 
divergence proportional to the one-loop color tensor
is
\begin{equation}
\mathcal{A}^{\oneloop}(1_{\phi},2_{\phi},3_{\phi},4_{\phi})\bigr|_{D=8\,\mathrm{div.}}
=\frac{i}{\epsilon}\frac{1}{(4\pi)^4}g^4\, b_1^{\oneloop} \, 
\frac{D_s+18}{120}(s^2+t^2+u^2)+\cdots\,.
\label{phiphiphiphiDivOneloopD8}
\end{equation}
For two pairs of distinct external scalars, we have have the gauge-theory 
divergence,
\begin{equation}
\mathcal{A}^{\oneloop}(1_{\phi_1},2_{\phi_1},3_{\phi_2},4_{\phi_2})
  \bigr|_{D=8\,\mathrm{div.}}
=\frac{i}{\epsilon}\frac{1}{(4\pi)^4}g^4b_1^{\oneloop}
  \frac{(D_s-2)s^2-40tu}{120} + \cdots\,,
\label{phi1phi1phi2phi2DivOneloopD8}
\end{equation}
while the two-scalar two-gluon divergence is
\begin{eqnarray}
\mathcal{A}^{\oneloop}(1_g,2_g,3_{\phi},4_{\phi})\bigr|_{D=8\,\mathrm{div.}}
&=&\frac{i}{\epsilon}\frac{1}{(4\pi)^4}g^4b_1^{\oneloop}\frac{1}{180}\Big[(D_s-2)s(2k_1\cdot\varepsilon_2k_2\cdot\varepsilon_1-\varepsilon_1\cdot\varepsilon_2\,s) 
 \nonumber \\
&& \hskip .2 cm \null 
+60(2k_3\cdot\varepsilon_1k_4\cdot\varepsilon_2\,t
 +2k_4\cdot\varepsilon_1k_3\cdot\varepsilon_2\,u
   +\varepsilon_1\cdot\varepsilon_2\,tu\Big]\, +\cdots\,. \nn \\
\label{AAphiphiDivOneloopD8}
\end{eqnarray}
In these eight-dimensional expressions the number of real scalars
circulating in the loops is $D_s - 8$.

As before, we obtain the corresponding 
half-maximal supergravity divergences with $n_\subV = D_s - 8$ matter multiplets by substituting 
the color factors in the Yang-Mills expressions with the kinematic numerator (\ref{N4Replacement}):
\begin{eqnarray}
&& \hskip -.4 cm 
\mathcal{M}^{\oneloop}(1_{\subV},2_{\subV},3_{\subV},4_{\subV})\bigr|_{D=8\,\mathrm{div.}}
= -\frac{1}{\epsilon}\frac{1}{(4\pi)^4}\Bigl(\frac{\kappa}{2}\Bigr)^4
stA_{Q=16}^{\tree}\frac{D_s+18}{120}(s^2+t^2+u^2)\,, \nonumber \\
&& \hskip -.4 cm 
\mathcal{M}^{\oneloop}(1_{\subV_1},2_{\subV_1},3_{\subV_2},4_{\subV_2})\bigr|_{D=8\,\mathrm{div.}}
= -\frac{1}{\epsilon}\frac{1}{(4\pi)^4}\Bigl(\frac{\kappa}{2}\Bigr)^4
s t A_{Q=16}^{\tree}\frac{(D_s-2)s^2-40tu}{120}\,,  \\
&& \hskip -.4 cm 
\mathcal{M}^{\oneloop}(1_{\subH},2_{\subH},3_{\subV},4_{\subV})\bigr|_{D=8\,\mathrm{div.}}
=-\frac{1}{\epsilon}\frac{1}{(4\pi)^4}\Bigl(\frac{\kappa}{2}\Bigr)^4
stA_{Q=16}^{\tree}\frac{1}{180} \nonumber\\
&& \hspace{5.5cm}\null\times 
\Bigl[(D_s-2)s(2k_1\cdot\varepsilon_2k_2\cdot\varepsilon_1-
   \varepsilon_1\cdot\varepsilon_2\,s)  \nonumber \\
& &\hspace{6.5cm}\null 
+60(2k_3\cdot\varepsilon_1k_4\cdot\varepsilon_2\,t
  +2k_4\cdot\varepsilon_1k_3\cdot\varepsilon_2\,u
   +\varepsilon_1\cdot\varepsilon_2\,tu)\Bigr]\,. \nonumber
\end{eqnarray}
Combined with the result in Ref.~\cite{HalfMax5D}, this gives the complete set of four-point divergences in $D=8$ for any
external states, whether in the graviton multiplet or in a vector
multiplet.

%%%%%%%%%%%%%%%%%%%%%%%%%%%%%%%%%%%%%%%%%%%%%%%%%%

\section{Two- and three-loop divergences}
\label{TwoLoopDivergenceSection}

In this section we systematically list out the two-loop four-point
divergences of half-maximal supergravity with external matter
multiplets in $D=4,5,6$ along with the divergences of corresponding
nonsupersymmetric gauge theory that control them. In all our
expressions we always subtract subdivergences.  As it turns out, in
$D=4$ the divergence appears to be of a form where it seemingly
could be an iteration of the one-loop divergence, so to conclusively
demonstrate that new divergences occur, we also present the three-loop
divergences.  Two-loop supergravity counterterms are provided in the
appendix.  As at one loop all divergences and counterterms in $D=4$
carry a manifest $SO(6)\times SO(n_\subV)$ symmetry.

\subsection{Four dimensions}

In $D=4$, renormalizability dictates that gauge-theory counterterms
for the four-gluon divergence and the two-gluon two-scalar divergence
contain only tree-level color tensors. Thus from simple
renormalizability considerations, we have~\cite{HalfMax5D}
\begin{eqnarray}
\mathcal{A}^{\twoloop}(1_g,2_g,3_g,4_g)\bigr|_{D=4\,\mathrm{div.}}&=& 0 + \cdots \,,
\nonumber \\
\mathcal{A}^{\twoloop}(1_g,2_g,3_{\phi},4_{\phi})\bigr|_{D=4\,\mathrm{div.}} &=& 0 
+ \cdots \,,
\label{TwoLoopDivRenormD4}
\end{eqnarray}
where ``+$\cdots$'' refers to dropped terms that contain tree and one-loop 
color tensors.  As noted in \sect{AmplitudesSubsection}, the dropped
terms are not needed for converting to supergravity.

As already discussed in Ref.~\cite{HalfMax5D},
renormalizability considerations do not protect four-scalar
divergences from containing higher-loop color tensors.
However, it turns out that when the four scalars are identical, the
divergences proportional to the two-loop color tensors cancel because
of a color identity.  This happens when all the scalars are identical 
because the two-loop
color tensors appear fully symmetrized in their indices, {\it i.e.} as
\begin{equation}
\phi^a \phi^b \phi^c \phi^d (c^{{\rm P} abcd}_{1234} + c_{3421}^{{\rm P} abcd}
 + c_{1423}^{{\rm P} abcd} + c_{2341}^{{\rm P} abcd} 
 + c_{1342}^{{\rm P} abcd} + c_{4231}^{{\rm P} abcd} ) = 0 + \cdots \,.
\label{TwoLoopDivSymmD4}
\end{equation}
By re-expressing these color factors in the basis
(\ref{TwoLoopColorBasis}), we immediately see that the two-loop color
tensors $b_1^{\twoloop}$ and $b_2^{\twoloop}$ cancel out, so there can be no
divergence containing these color tensors when the four scalars are
identical:
\begin{equation}
\mathcal{A}^{\twoloop}(1_{\phi},2_{\phi},3_{\phi},4_{\phi})\bigr|_{D=4\,\mathrm{div.}}
= 0 + \cdots \,.
\label{eq:2loopDEq4YM}
\end{equation}

%%%%%%%%%%%%%%%%%%%%
%FIGURE
%
\begin{figure}
\begin{center}
\includegraphics[scale=0.5]{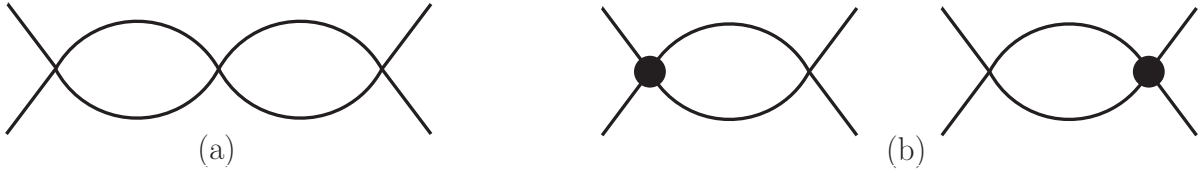}
\caption{The divergences containing a two-loop color tensor and a
  factor of $D_s^2$ all arise from diagrams of the form (a).  The
  diagrams subtracting the one-loop subdivergences are shown in (b). }
\label{Ds2TermsFigure}
\end{center}
\end{figure}
%%%%%%%%%%%%%%%%%%%%%%

If the scalars are not all identical, the previous symmetry argument
no longer applies. Indeed, we find that gauge-theory amplitudes with
non-identical external scalars are divergent.  The nonvanishing
contribution to the two-loop four-scalar divergence with a distinct
pair of scalars that contains an independent two-loop color tensor is
\begin{equation}
\mathcal{A}^{\twoloop}(1_{\phi_1},2_{\phi_1},3_{\phi_2},4_{\phi_2})
\bigr|_{D=4\,\mathrm{div.}}
=\frac{i}{\epsilon^2}\frac{1}{(4\pi)^4}g^6b_1^{\twoloop}\frac{(D_s-2)^2}{4} + \cdots\,,
\label{TwoloopDEq4YMDiv}
\end{equation}
where the one-loop subdivergences have all been subtracted.  Since the
number of scalars is $D_s -4$, the divergence
does not vanish for any (positive) number of scalar fields.  In
this case the $t$-channel basis color tensor $b_2^{\twoloop}$ is absent.

A curious feature of the divergence in \eqn{TwoloopDEq4YMDiv} is that
it does not contain a $1/\eps$ divergence but only a $1/\eps^2$
divergence.  This may be understood straightforwardly for terms
proportional to the square of the state-counting parameter $D_s$.
Prior to subtractions, the only diagrams that give contributions that
contain both a factor of $D_s^2$ and a two-loop color tensor are those
of the form in \fig{Ds2TermsFigure}(a).  Without this configuration it
is not possible to get two factors of $D_s$, each of which come from
contracting a Lorentz index around an independent loop.  A second
source of these terms is the subtraction diagrams where one factor
of $D_s$ comes from the one-loop
subtraction  and the second comes from a
loop, as illustrated in \fig{Ds2TermsFigure}(b).  The structure of the
result follows from the fact that each of the contributing loops is of the
form,
\begin{equation}
 \frac{a}{\eps} + b \,,
\end{equation}
where $a$ and $b$ are parameters that depend on the external momenta.
The two-loop diagram (\fig{Ds2TermsFigure}(a)) contains two such loops and is just the 
square of this:
\begin{equation}
V^{\rm (a)} = \Bigl(\frac{a}{\eps} + b \Bigr) \Bigl(\frac{a}{\eps} + b\Bigr) \,,
\label{eq:direct}
\end{equation}
times a prefactor.  The subtraction terms in \fig{Ds2TermsFigure}(b) are  
of the form,
\begin{equation}
V^{\rm (b)} = -\frac{a}{\eps} \Bigl(\frac{a}{\eps} + b\Bigr) 
      - \Bigl(\frac{a}{\eps} + b\Bigr) \frac{a}{\eps} \,,
\label{eq:subtraction}
\end{equation}
times the same prefactor.  Combining the direct terms~\eqref{eq:direct} with the
subtraction terms~\eqref{eq:subtraction} 
flips the sign of the $1/\eps^2$ terms and cancels
the $1/\eps$ terms, as given in \eqn{TwoloopDEq4YMDiv}.  The
terms that are subleading in $D_s$ are more complicated because
other diagrams contribute.  Once all pieces are added together, we find that all
$1/\eps$ terms cancel for the divergence, as for the $D_s^2$
terms. Since the supergravity divergence is inherited from the
gauge-theory one, it too will not have $1/\eps$ contributions.

Converting the lack of gauge-theory divergences in 
\eqns{TwoLoopDivRenormD4}{eq:2loopDEq4YM}
to supergravity divergences using the double-copy 
substitution in \eqn{TwoLoopSubstitution} gives us
the following finiteness results:
\begin{eqnarray}
&& \mathcal{M}^{\twoloop}(1_{\subH},2_{\subH},3_{\subH},4_{\subH})\bigr|_{D=4\,\mathrm{div.}} = 0\,,  \nonumber \\
&& \mathcal{M}^{\twoloop}(1_{\subH},2_{\subH},3_{\subV},4_{\subV})\bigr|_{D=4\,\mathrm{div.}} = 0\,, \nonumber \\
&& \mathcal{M}^{\twoloop}(1_{\subV},2_{\subV},3_{\subV},4_{\subV})\bigr|_{D=4\,\mathrm{div.}} = 0 \,.
\end{eqnarray}
These hold for all external states in the respective multiplets,
independent of the number of matter multiplets added to the theory.
The vanishing of the divergences in the four
identical-matter-multiplet case can also be seen from its symmetry.
From dimensional analysis, after extracting the crossing-symmetric
factor of $stA_{Q=16}^\tree$, there is an additional factor of $s$
which can appear only in the crossing symmetric form $s+t+u
=0$, implying the vanishing of the divergence.  We can think of this
cancellation as ``accidental,'' similar to the vanishing of one-loop
divergences in pure Einstein gravity.  Interestingly, all the above
finite results exhibit cancellation separately in both the
unsubtracted pieces and in the subtractions when a uniform mass
infrared regulator is used.  This is true even when one-loop
divergences imply the presence of subdivergences in higher-loop
amplitudes.  Another notable case where this happens is three-loop
$\mathcal{N}=4$ supergravity in four dimensions with internal
matter~\cite{N4gravThreeLoops}.

Finally, applying the substitution rule (\ref{TwoLoopSubstitution}) to the 
gauge-theory amplitude with a pair of distinct scalars gives 
the nonvanishing divergence for a four-point supergravity 
amplitude with external states from a pair of distinct vector multiplets,
\begin{equation}
\mathcal{M}^{\twoloop}(1_{\subV_1},2_{\subV_1},3_{\subV_2},4_{\subV_2})\bigl|_{D=4\,\mathrm{div.}}
=-\frac{1}{\epsilon^2}\frac{1}{(4\pi)^4}
\Bigl(\frac{\kappa}{2}\Bigr)^6s^2tA_{Q=16}^{\tree}\frac{(D_s-2)^2}{4}\,,
\label{TwoloopDEq4GravDiv}
\end{equation}
where the state-counting parameter takes the value, $D_s = n_\subV +
D$, with $D=4$ in the four-dimensional helicity scheme~\cite{FDH}.
Since the divergence is for a pair of distinct matter multiplets, for
consistency we need $n_\subV \ge 2$. 

The fact that in $D=4$ the two-loop divergence
(\ref{TwoloopDEq4GravDiv}) looks similar to the result from
iterating the one-loop divergences raises the question\footnote{We
  thank G. Bossard, K. Stelle and P. Howe for raising this question.}
of whether it might follow from the one-loop divergences in
\eqn{eq:DEq4Grav}.  To
definitively settle any such potential question, we have also
computed the complete set of three-loop $\mathcal{N}=4$ supergravity
divergences in $D=4$.  Our results for these divergences are
\begin{eqnarray}
\mathcal{M}^{(3)}(1_\subH,2_\subH,3_\subH,4_\subH)\bigr|_{D=4\,\mathrm{div.}} & = & 0 \,, 
 \nonumber\\
\mathcal{M}^{(3)}(1_\subH,2_\subH,3_\subV,4_\subV)\bigr|_{D=4\,\mathrm{div.}} & = & 0 \,,
 \nonumber \\
\mathcal{M}^{(3)}(1_\subV,2_\subV,3_\subV,4_\subV)\bigr|_{D=4\,\mathrm{div.}}
&=&-\frac{1}{(4\pi)^6}\Bigl(\frac{\kappa}{2}\Bigr)^8
           (s^2+t^2+u^2)stA_{Q=16}^{(0)} 
\nonumber\\
&&\hspace{2.5cm}
\null \times\frac{(D_s-2)^2}{4}\biggl(\frac{D_s-2}{2\epsilon^3}
  -\frac{1}{\epsilon^2} +\frac{1}{\epsilon}\biggr)\,, \nonumber \\
\mathcal{M}^{(3)}(1_{\subV_1},2_{\subV_1},3_{\subV_2},4_{\subV_2})
  \bigr|_{D=4\,\mathrm{div.}} &=&
   -\frac{1}{(4\pi)^6}\left(\frac{\kappa}{2}\right)^8
   stA_{Q=16}^{(0)}
\label{ThreeLoopDivergences} \\
&&\times\biggl[\frac{1}{\epsilon^3}\frac{(D_s-2)^2}{8}
\left((D_s-4)s^2-4tu\right)+\frac{1}{\epsilon^2}\frac{(D_s-2)^2}{2}tu \nonumber \\
&& 
\null +\frac{1}{\epsilon}\biggl(\frac{(D_s-2)^2}{2}(s^2+tu)-(7D_s-38)(s^2+2tu)\zeta_3\biggr)\biggr]\,, \hskip .5 cm  \nonumber
\end{eqnarray}
where the number of vector multiplets is $n_\subV = D_s - 4$ and all
subdivergences have been subtracted, as usual.  The vanishing of
divergences when all four external states are from the graviton
multiplet was shown in Ref.~\cite{N4gravThreeLoops}.  The vanishing of
divergences when two external states are from the graviton multiplet
and two from a matter multiplet is new.  We leave the
comparison of these divergences to those of nonsupersymmetric
Yang-Mills coupled to scalars to future studies.

The divergences (\ref{ThreeLoopDivergences}) are not of a
form where they can be induced by the one-loop divergences, settling 
any potential issues on whether these are new divergences.  In
particular, we cannot obtain a $\zeta_3$ from a one-loop divergence.
If we ignore, for the moment, any potential issues with the
$SL(2,\mathbb{R})$ duality anomaly, this result contradicts the
expected finiteness~\cite{BHSNew}, had an off-shell superspace
manifesting all 16 supercharges existed~\cite{BossardHoweStelle5D}.

Now consider the $SL(2,\mathbb{R})$ duality anomaly.  One may wonder
if it can somehow be responsible for the divergences in
\eqns{TwoloopDEq4GravDiv}{ThreeLoopDivergences}, since it can prevent
use of the duality symmetry to rule out a counterterm~\cite{BHSNew}.
However, these divergences are not of the proper form had they been due
to the anomaly.  Anomalies are associated with a ``0/0,'' or more precisely
in dimensional regularization, contributions of ${\cal O}(\eps)$ that
violate a symmetry and can give an ${\cal O}(\eps^0)$ contribution
when they hit a $1/\eps$ divergence.  Indeed, this is how the
anomaly enters into one-loop amplitudes~\cite{RaduAnomaly}.  At two
loops these finite one-loop terms could feed in to give at most a
$1/\eps$ divergence, and we would have found that at two loops the
divergences would contain no $1/\eps^2$ term and at three loops no
$1/\eps^3$ terms.  In addition the amplitudes containing the
divergences are all inert under the anomalous $U(1)$, using the
helicity counting rules of Ref.~\cite{RaduAnomaly}.  These features
are incompatible with the anomaly being the source of the $D=4$ divergences.
We therefore conclude that our results are inconsistent with the
existence of a 16-supercharge off-shell superspace in $D=4$.

%%%%%%%%%%%%%%%%%
\subsection{Five dimensions}

The two-loop gluon amplitudes of five-dimensional gauge theory 
coupled to scalars have divergences due to an $F^3$ operator. This
operator generates divergences containing only tree-level color 
factors, and hence no two-loop color tensors are present~\cite{HalfMax5D}.
Thus, the four-gluon divergence is given by
\begin{equation}
\mathcal{A}^{\twoloop}(1_g,2_g,3_g,4_g)\bigr|_{D=5\,\mathrm{div.,}}=
0 + \cdots \,.
\end{equation}
However, when we have external adjoint scalars, counterterms involving the two-loop color tensors exist.  The available counterterms
involving two-loop color tensors at two loops in five dimensions are
similar to those involving one-loop color tensors at one loop in six
dimensions.  For four external scalars, we have $D^2\phi^4$.  For
identical scalars, the two-loop gauge-theory divergence is
\begin{equation}
\mathcal{A}^{\twoloop}(1_{\phi},2_{\phi},3_{\phi},4_{\phi})\bigr|_{D=5\,\mathrm{div.,}}
=\frac{i}{\epsilon}\frac{1}{(4\pi)^5}
g^6\frac{(10-D_s)\pi}{3}\left(b_1^{\twoloop}(u-s)+b_2^{\twoloop}(u-t)\right)+\cdots\,.
\end{equation}
For distinct external scalars, the divergence containing two-loop color tensors is
\begin{equation}
\mathcal{A}^{\twoloop}(1_{\phi_1},2_{\phi_1},3_{\phi_2},4_{\phi_2})
\bigr|_{D=5\,\mathrm{div.}}
=\frac{i}{\epsilon}\frac{1}{(4\pi)^5}g^6\frac{(10-D_s)\pi}{6}
\left(b_1^{\twoloop}(t-3s)+b_2^{\twoloop}(t-u)\right) + \cdots\,.
\end{equation}
Finally for the two-scalar two-gluon divergence,
we have a $\phi^2F^2$ counterterm.  The divergence corresponding
to this operator is
\begin{equation}
\mathcal{A}^{\twoloop}(1_g,2_g,3_{\phi},4_{\phi})\bigr|_{D=5\,\mathrm{div.}}
=\frac{i}{\epsilon}\frac{1}{(4\pi)^5}g^6b_1^{\twoloop}
\frac{(10-D_s)\pi}{6}\left(\varepsilon_1\cdot\varepsilon_2\,s
       - 2k_1\cdot\varepsilon_2\,k_2\cdot\varepsilon_1\right)+\cdots\,.
\end{equation}

As in $D=4$, we can convert these results to those of half-maximal
supergravity by replacing the color tensors with the kinematic
numerators in \eqn{TwoLoopSubstitution} to yield the supergravity divergences:
\begin{eqnarray}
&& \mathcal{M}^{\twoloop}(1_{\subH},2_{\subH},3_{\subH},4_{\subH})\bigr|_{D=5\,\mathrm{div.}}
= 0\,, \nonumber \\
&& \mathcal{M}^{\twoloop}(1_{\subV},2_{\subV},3_{\subV},4_{\subV})\bigr|_{D=5\,\mathrm{div.}}
=\frac{1}{\epsilon}\frac{1}{(4\pi)^5}\Bigl(\frac{\kappa}{2}\Bigr)^6
stA_{Q=16}^{\tree}\frac{(10-D_s)\pi}{3}(s^2+t^2+u^2)\,, \nonumber \\
&& \mathcal{M}^{\twoloop}(1_{\subV_1},2_{\subV_1},3_{\subV_2},4_{\subV_2})
\bigr|_{D=5\,\mathrm{div.}}
=\frac{1}{\epsilon}\frac{1}{(4\pi)^5}\Bigl(\frac{\kappa}{2}\Bigr)^6
 stA_{Q=16}^{\tree}\frac{(10-D_s)\pi}{6}(3s^2+2tu)\,, \nonumber \hskip 1.5 cm \\
&& \mathcal{M}^{\twoloop}(1_{\subH},2_{\subH},3_{\subV},4_{\subV})
 \bigr|_{D=5\,\mathrm{div.}}
=-\frac{1}{\epsilon}\frac{1}{(4\pi)^5}\Bigl(\frac{\kappa}{2}\Bigr)^6
s^2tA_{Q=16}^{\tree}\frac{(10-D_s)\pi}{6} \nn \\
&& \hskip 8 cm \null \times
\left(\varepsilon_1\cdot\varepsilon_2\,s
 -2k_1\cdot\varepsilon_2\,k_2\cdot\varepsilon_1\right)\,, 
\label{TwoLoopSugraDiv}
\end{eqnarray}
where the number of matter multiplets is $n_{\subV} = D_s - 5$.  An
interesting feature of these divergences is that they vanish if the
number of vector supermultiplets is $n_{\subV}= 5$, which corresponds 
to the theory of $\mathcal{N}=1$, $D=10$ supergravity dimensionally reduced to 
$D=5$. It would be interesting to know if these cancellations
for five vector multiplets persist to higher-loop orders.  It
is noteworthy that a linearized superspace exists for this
theory in $D=10$~\cite{HoweSuperspace}.

The divergent results in \eqn{TwoLoopSugraDiv} are in direct conflict
with the predictions of Ref.~\cite{BHSNew}, under the
assumption~\cite{BossardHoweStelle5D} that there is a harmonic
superspace manifesting all 16 supercharges off shell.   On the other hand,
the two-loop ultraviolet finiteness of $D=5$ pure half-maximal
supergravity is a direct consequence of the duality between color and
kinematics and the general structure of corresponding gauge-theory
divergences, so from this vantage point there is no mystery. 

\subsection{Six dimensions}

Half-maximal supergravity is divergent in six dimensions with or
without matter in the loop~\cite{HalfMax5D}.  This can be understood
through the presence of an $F^4$ counterterm in pure Yang-Mills theory that
contains the independent two-loop color tensors.  The divergence was
given in Ref.~\cite{HalfMax5D}, so we do not reproduce it here.  One
fact of interest is that there is no $1/\epsilon^2$ term in the
divergence for pure half-maximal supergravity (no matter in the loop),
which is consistent with expectations based on the lack of 
one-loop divergences in pure half-maximal supergravity in six dimensions.

The four-point Yang-Mills divergence for two scalars and two gluons is
given by a $D^2\phi^2F^2$ operator.  We do not give the divergence or
the associated two-external-matter four-point supergravity divergence,
but we do mention the presence of a factor of $26-D_s$ multiplying the
$1/\epsilon^2$ piece in both.  This is again consistent with the one-loop
subdivergence.

For four external scalars, counterterms of the form $D^4\phi^4$ are
valid.  The corresponding divergence involving the two-loop
color tensors is given by
\begin{eqnarray}
\mathcal{A}^{\twoloop}(1_{\phi},2_{\phi},3_{\phi},4_{\phi})\bigr|_{D=6\,\mathrm{div.}}&=&i\frac{1}{(4\pi)^6}g^6\frac{1}{144}\left(\frac{(D_s-6)(26-D_s)}{\epsilon^2}-\frac{13D_s+142}{3\epsilon}\right) \nonumber \\
&&\hspace{2.5cm}\times\left(b_1^{\twoloop}t(s-u)
 +b_2^{\twoloop}s(t-u)\right) + \cdots \,.
\hskip 1.2 cm 
\end{eqnarray}
 We do not present the Yang-Mills divergence for two different scalars
 because it is somewhat complicated and not particularly enlightening.
 We do note that once again a factor of $26-D_s$ in the $1/\epsilon^2$
 pieces multiplies each color tensor.

Applying the substitution rule (\ref{TwoLoopSubstitution})
to the gauge-theory results, we have for half-maximal supergravity divergences with external matter states:
\begin{eqnarray}
\mathcal{M}^{\twoloop}(1_{\subV},2_{\subV},3_{\subV},4_{\subV})\bigr|_{D=6\,\mathrm{div.}}
&=&\frac{1}{(4\pi)^6}\Bigl(\frac{\kappa}{2}\Bigr)^6s^2t^2u
A_{Q=16}^{\tree} \nn \\
&&\hspace{1.5cm}\times \null
\frac{1}{48}\left(\frac{(D_s-6)(26-D_s)}{\epsilon^2}
-\frac{13D_s+142}{3\epsilon}\right)\,, \nonumber \\
\mathcal{M}^{\twoloop}(1_{\subV_1},2_{\subV_1},3_{\subV_2},4_{\subV_2})\bigr|_{D=6\,\mathrm{div.}}
&=&\frac{1}{(4\pi)^6}\Bigl(\frac{\kappa}{2}\Bigr)^6s^2 t 
   A_{Q=16}^{\tree} \nonumber \\
&&\times\frac{1}{144}\left[\frac{26-D_s}{\epsilon^2}\left((D_s-6)s^2-5(s^2+t^2+u^2)\right)\right. \nonumber \\
&&\hspace{.5cm}\left.-\frac{1}{3\epsilon}\left((13D_s+142)s^2+\frac{13D_s-578}{2}(s^2+t^2+u^2)\right)\right]\,. \nn \\
\end{eqnarray}
As always, the result when the four external states are from a single
vector multiplet can be obtained from the result when the states are
from a distinct pair of vector multiplets simply by summing over the
independent external permutations.  The complete set of counterterms
for external gravitons and matter vectors may be found in
\App{CounterTermAppendix}.

%%%%%%%%%%%%%%%%%%%%%%%%%%%%%%%%%%%%%%%%%%%%%%%%%%%
\section{Conclusions and Outlook}
\label{ConclusionSection}

%%%%%%%%%%%% TABLE %%%%%%%%%%%%%%%%%%%%%%%%%%
\begin{table*}[t]\caption{A schematic table of the
counterterms of half-maximal supergravity with matter multiplets in various
dimensions at one and two loops, together with corresponding gauge-theory
counterterms with the appropriate color tensors. The displayed supergravity
counterterms are for gravitons and vector matter multiplets. For the
gauge-theory case they are for the gluons and scalars of the theory.
\label{OperatorTable} }
\vskip .4 cm
\begin{center}
\begin{tabular}{||l||C{1.8cm}|C{1.8cm}|C{1.8cm}||C{1.8cm}|C{1.8cm}|C{1.8cm}||}
\hline
\multicolumn{1}{||c||}{{\multirow{2}{*}{\large{Amplitude}}}} & \multicolumn{3}{|c||}{One Loop} & \multicolumn{3}{|c||}{Two Loops} \\
\cline{2-7}
 & $D=4$ & $D=6$ & $D=8$ & $D=4$ & $D=5$ & $D=6$  \\
\hline
\hline
$\mathcal{A}^{(L)}(1_g,2_g,3_g,4_g)$ & finite & finite & $F^4$ & finite & finite & $F^4$ \\
\hline
$\mathcal{M}^{(L)}(1_\subH,2_\subH,3_\subH,4_\subH)$ & finite & finite & $R^4$ & finite & finite & $D^2R^4$ \\
\hline
\hline
$\mathcal{A}^{(L)}(1_g,2_g,3_{\phi},4_{\phi})$ & finite & $\phi^2F^2$ & $D^2\phi^2F^2$ & finite & $\phi^2F^2$ & $D^2\phi^2F^2$ \\
\hline
$\mathcal{M}^{(L)}(1_\subH,2_\subH,3_\subV,4_\subV)$ & finite & $F^2R^2$ & $D^2F^2R^2$ & finite & $D^2F^2R^2$ & $D^4F^2R^2$ \\
\hline
\hline
$\mathcal{A}^{(L)}(1_{\phi},2_{\phi},3_{\phi},4_{\phi})$ & $\phi^4$ & finite & $D^4\phi^4$ & finite & $D^2\phi^4$ & $D^4\phi^4$ \\
\hline
$\mathcal{M}^{(L)}(1_\subV,2_\subV,3_\subV,4_\subV)$ & $F^4$ & finite & $D^4F^4$ & finite & $D^4F^4$ & $D^6F^4$ \\
\hline
\hline
$\mathcal{A}^{(L)}(1_{\phi_1},2_{\phi_1},3_{\phi_2},4_{\phi_2})$ & $\phi^4$ & $D^2\phi^4$ & $D^4\phi^4$ & $\phi^4$ & $D^2\phi^4$ & $D^4\phi^4$ \\
\hline
$\mathcal{M}^{(L)}(1_{\subV_1},2_{\subV_1},3_{\subV_2},4_{\subV_2})$ & $F^4$ & $D^2F^4$ & $D^4F^4$ & $D^2F^4$ & $D^4F^4$ & $D^6F^4$ \\
\hline
\end{tabular}
\end{center}
\end{table*}
%%%%%%%%%%%%%%%%%%%%%%%%%%%%%%%%%%%%%%%%%%%%%%%%%

In this paper we mapped out the one- and two-loop four-point
divergences of half-maximal supergravity including abelian-vector
matter multiplets in various dimensions.  In particular, we showed
that half-maximal supergravity with matter multiplets does contain new
two-loop ultraviolet divergences in $D=4,5,6$.  We also worked out the
four-point divergences at three loops in $D=4$ to conclusively show
that new divergences do occur in $D=4$.  The $D=4$ theory has long
been known to be divergent at one loop~\cite{Fischler,Arkady}, which
we confirmed here as well.  Our one- and two-loop results are
summarized in \tab{OperatorTable}, which shows a schematic form of the
counterterms of half-maximal supergravity, as well as those of
nonsupersymmetric gauge theory involving the color tensors that
control the gravity divergences.  (We do not include odd dimensions at
one loop because those are automatically finite when using dimensional
regularization.)

Bossard, Howe and Stelle recently
conjectured~\cite{BossardHoweStelle5D} the existence of 16-supercharge
linearly realized harmonic superspaces in $D=4$ and in $D=5$ in order
to explain the finiteness of pure half-maximal supergravity at three
loops in $D=4$ and at two loops in
$D=5$~\cite{N4gravThreeLoops,VanhoveN4,HalfMax5D}.  Such superspaces
have the appealing feature that no new ``miracles'' would be required
to explain the observed finiteness.  Very recently they
argued~\cite{BHSNew} that if the conjectured superspace were to exist
in $D=5$, then there would be no new two-loop divergences even when
matter multiplets are added to the theory.  In $D=4$ the situation is
similar except for the appearance of an anomaly~\cite{MarcusAnomaly}
in the rigid $SL(2,\mathbb{R})$ duality symmetry.  However, we found
that the calculated divergences in $D=4$ are not compatible with them
being due to the anomaly.  The results of the present paper then show
that there are new divergences in all these cases, contradicting the
predictions had the desired superspaces existed in $D=4$ and $D=5$.

We emphasize that there is no mystery in the half-maximal supergravity
divergence structure from the vantage point of the duality between
color and kinematics.  At one and two loops, it shows in a direct way
why amplitudes with external matter can diverge when the purely
external-graviton-multiplet case does not, linked to the well
understood divergences of nonsupersymmetric gauge theory.  In
addition, it gives us the means to precisely determine the
coefficients of the divergences.

The one- and two-loop cases analyzed in this paper are especially
simple because the maximal super-Yang-Mills numerators used in the
double-copy construction are independent of loop momenta.  For higher
loops the situation is more complex to analyze because loop momenta
enter into the super-Yang-Mills numerators, altering the form of the
integrals compared to those of nonsupersymmetric gauge theory.
Nevertheless, as suggested in Ref.~\cite{HalfMax5D}, we expect the
divergences of half-maximal supergravity to be related to the
divergence structure of corresponding nonsupersymmetric gauge-theory
amplitudes.  We look forward to new calculations that will shed further
light on the origin of the remarkably good ultraviolet behavior of
pure supergravity theories with 16 or more supercharges.

\vskip .2 cm 
\subsection*{Acknowledgments}
We thank G.~Bossard, P.~Howe and K.~Stelle for explaining their work
to us prior to publication of their paper and for motivating us to
study the divergences in half-maximal supergravity with matter
multiplets.  We also thank them for many helpful discussions.  We also thank
J.~J.~M.~Carrasco, H.~Johansson, J.~Nohle and R.~Roiban for many
helpful discussions and comments on the manuscript.  This research was
supported by the US Department of Energy under contract
DE-FG02-13ER42022. S.~D. was supported by a US Department of Energy
Graduate Student Fellowship under contract DE-SC0008279.

%\newpage
\appendix

\section{Supergravity counterterms}
\label{CounterTermAppendix}

Any potential symmetry explanation of the vanishings of potential
divergences of supergravity must also properly give the allowed
counterterms in detail.  For this purpose, in this appendix, we give
counterterms corresponding to our calculated one- and two-loop
divergences in half-maximal supergravity coupled to $n_{\subV}=D_s-D$
matter vector multiplets in various dimensions.  For each divergence,
we give a counterterm for a particular field content, focusing on
external graviton and abelian-vector matter states.  These can in turn
be supersymmetrized using the supersymmetric form of the divergences
given \sects{OneLoopDivergenceSection}{TwoLoopDivergenceSection}, but
we do not do so here.  The fact that this theory diverges at one loop
in $D=4$ has been known since the early days of
supergravity~\cite{Fischler,Arkady}.  Here we map out the full set of
counterterms for four external gravitons or abelian-vector matter
states at one and two loops.

For notational simplicity, we define contractions of field strengths as
\begin{equation}
(FF)\equiv F_{\mu\nu}F^{\mu\nu}, \hspace{1cm} (FFFF)\equiv F_{\mu\nu}F^{\nu\rho}F_{\rho\sigma}F^{\sigma\mu}\,.
\label{eq:appNotation}
\end{equation}
We also allow derivatives in this notation; for example,
\begin{equation}
(D_{\alpha\beta}FFD^{\alpha}FD^{\beta}F)\equiv (D_{\alpha}D_{\beta}F_{\mu\nu})F^{\nu\rho}(D^{\alpha}F_{\rho\sigma})(D^{\beta}F^{\sigma\mu})\,,
\end{equation}
where we have also introduced the notation, $D_{\mu\nu}\equiv
D_{\mu}D_{\nu}$.  For $F^2R^2$ type operators, when no indices are
written, the first two indices of one Riemann tensor are understood
to be contracted with the first two indices of the other Riemann
tensor, while the last two indices of each obey the relations of the field
strengths in Eq.~\eqref{eq:appNotation}:
\begin{eqnarray}
&&(RR)\equiv R_{\mu\nu\lambda\gamma}R^{\mu\nu\lambda\gamma},\hspace{1cm} (RF)(RF)\equiv R_{\mu\nu\lambda\gamma}F^{\lambda\gamma}R^{\mu\nu}_{\hphantom{\mu\nu}\delta\kappa}F^{\delta\kappa}, \nn \\
&&\hspace{2.5cm}(RFRF)\equiv R_{\mu\nu\lambda\gamma}F^{\gamma\delta}R^{\mu\nu}_{\hphantom{\mu\nu}\delta\kappa}F^{\kappa\lambda}\,.
\end{eqnarray}
In cases where the first two indices of a Riemann tensor are not contracted with the other one, we will write them out explicitly:
\begin{equation}
(R_{\mu\nu}R^{\mu}_{\hphantom{\mu}\rho}D^{\nu}FD^{\rho}F)\equiv R_{\mu\nu\lambda\gamma}R^{\mu\hphantom{\rho}\gamma\delta}_{\hphantom{\mu}\rho}(D^{\nu}F_{\delta\kappa})(D^{\rho}F^{\kappa\lambda})\,.
\end{equation}
All $R$'s in the counterterm operators refer to Riemann tensors and
should not be confused with the Ricci tensor or Ricci scalar, despite
the notation.  The indices in $R^4$-type operators are written out
explicitly with the understanding that derivatives act only on the
tensor that they immediately precede, \emph{e.g.}
$D_{\alpha}R_{\mu\nu\lambda\gamma}R^{\mu\nu\lambda\gamma}
 \equiv(D_{\alpha}R_{\mu\nu\lambda\gamma})R^{\mu\nu\lambda\gamma}$.

Since the duality-satisfying numerators of maximally supersymmetric
Yang-Mills amplitudes are independent of loop momenta at one and two loops, we
exploit the double-copy property to construct our counterterm
operators, as was done in Ref.~\cite{HalfMax5D}.  The four-point
one-loop BCJ numerator for maximal super-Yang-Mills theory is given 
by a contraction of field strengths,
\begin{equation}
F^4=-2\biggl[(F_1F_2F_3F_4)-\frac{1}{4}(F_1F_2)(F_3F_4)
+ \mathrm{cyclic}(2,3,4)\biggr]\,,
\label{eq:oneLoopF4}
\end{equation}
while the two-loop numerator $s^2tA_{Q=16}^{\mathrm{tree}}$ is given by
\begin{eqnarray}
\null\hspace{-.9cm}D^2F^4&=&4\biggl[(D_{\alpha}F_1D^{\alpha}F_2F_3F_4)
+(D_{\alpha}F_1F_3F_4D^{\alpha}F_2)
+(D_{\alpha}F_1F_4D^{\alpha}F_2F_3)\vphantom{\frac{1}{4}} \nn \\
&&\null
-\frac{1}{4}(D_{\alpha}F_1D^{\alpha}F_2)(F_3F_4)
-\frac{1}{4}(D_{\alpha}F_1F_3)(D^{\alpha}F_2F_4)
-\frac{1}{4}(D_{\alpha}F_1F_4)(D^{\alpha}F_2F_3)\biggr]\,,
\end{eqnarray}
where the labels on the field strengths in these cases indicate the
corresponding external legs.  We use these expressions as replacements
for the color factors in operators generating the nonsupersymmetric
Yang-Mills divergences.  We then associate products of Yang-Mills
objects with gravity objects:
\begin{equation}
\phi_iF_{i\,\mu\nu}\rightarrow F_{i\,\mu\nu}, \hspace{1cm} F_{i\,\mu\nu}F_{i\,\rho\sigma}\rightarrow -2R_{i\,\mu\nu\rho\sigma}\,.
\label{eq:opSubs}
\end{equation}
At the linearized level, the products of Yang-Mills objects and the
gravity object each have the same contribution to the amplitude (see
Ref.~\cite{HalfMax5D} for more detail).

For example, the nonsupersymmetric Yang-Mills divergence for four
identical scalars at one loop in four dimensions involving the
one-loop color tensor \eqref{phi4OneLoop} is generated by
\begin{equation}
\frac{1}{\epsilon}\frac{1}{(4\pi)^2}g^4\frac{3(D_s-2)}{2}\frac{1}{4!}
\, b_1^{\oneloop abcd}\phi^a \phi^b \phi^c \phi^d\,.
\end{equation}
Substituting \eqn{eq:oneLoopF4} for $b_1^{\oneloop}$ and using
\eqn{eq:opSubs}, we have for the
operator generating the single-matter-vector divergence in
half-maximal supergravity,
\begin{equation}
\frac{1}{\epsilon}\frac{1}{(4\pi)^2}\Bigl(\frac{\kappa}{2}\Bigr)^4
  \frac{3(D_s-2)}{8}\biggl((FFFF)-\frac{1}{4}(FF)^2\biggr)\,.
\end{equation}
We will not provide the input Yang-Mills operators as they do not
generate the full divergences, but only generate the pieces proportional to the
color tensors of interest.  Nevertheless, the double-copy construction
is evident in the contraction structure of the indices in the gravity
counterterms, where Lorentz indices can be separated according to the
gauge theory to which they belong.  For each dimension and loop order, we
provide counterterms for the following four-point half-maximal
supergravity amplitudes:
\begin{itemize}
 \itemsep0em 
\item four external gravitons with matter included in the loop,
\item two external gravitons and two external vector matter states,
\item four external vector matter states belonging to the same multiplet,
\item four external vector matter states belonging to two different
  multiplets.  In this case the subscript labels on the field strengths
  indicate the matter multiplet to which the vector state belongs; these expressions are also valid for $i=j$, returning the counterterm for a single multiplet up to terms that vanish on shell.
\end{itemize}

\subsection{One Loop}

\subsubsection{Four Dimensions}
\begin{eqnarray}
&&\hspace{-.5cm}C^{\oneloop,\,D=4}_{(R,R,R,R)}=0\,, \nn \\
&&\hspace{-.5cm}C^{\oneloop,\,D=4}_{(R,R,F,F)}=0\,, \nn \\
&&\hspace{-.5cm}C^{\oneloop,\,D=4}_{(F,F,F,F)}=-\frac{1}{\epsilon}\frac{1}{(4\pi)^2}
\Bigl(\frac{\kappa}{2}\Bigr)^4\frac{3(D_s-2)}{8}\biggl((FFFF)
         -\frac{1}{4}(FF)^2\biggr)\,, \notag \\
&&\hspace{-.5cm}C^{\oneloop,\,D=4}_{(F_i,F_i,F_j,F_j)}=-\frac{1}{\epsilon}\frac{1}{(4\pi)^2}\Bigl(\frac{\kappa}{2}\Bigr)^4\frac{D_s-2}{4}\biggl((F_iF_iF_jF_j)+\frac{1}{2}(F_iF_jF_iF_j)\nn \\
&&\hspace{7.5cm} 
  -\frac{1}{8}(F_iF_i)(F_jF_j)-\frac{1}{4}(F_iF_j)^2\biggr)\,.
\end{eqnarray}

\subsubsection{Six Dimensions}
\begin{eqnarray}
&&C^{\oneloop,\,D=6}_{(R,R,R,R)}=0\,, \nn \\
&&C^{\oneloop,\,D=6}_{(R,R,F,F)}=\frac{1}{\epsilon}\frac{1}{(4\pi)^3}
\Bigl(\frac{\kappa}{2}\Bigr)^2 \, \frac{26-D_s}{24}
\biggl((RRFF)+\frac{1}{2}(RFRF) \nn \\
&&\hspace{8cm}-\frac{1}{8}(RR)(FF)-\frac{1}{4}(RF)(RF)\biggr)\,, \nn \\
&&C^{\oneloop,\,D=6}_{(F,F,F,F)}=0\,, \\
&&C^{\oneloop,\,D=6}_{(F_i,F_i,F_j,F_j)}=-\frac{1}{\epsilon}\frac{1}{(4\pi)^3}
\Bigl(\frac{\kappa}{2}\Bigr)^4\, \frac{26-D_s}{12}
\biggl((D_{\alpha}F_iD^{\alpha}F_iF_jF_j)+\frac{1}{2}(D_{\alpha}F_iF_jD^{\alpha}F_iF_j) \nn\\
&&\hspace{6.5cm}-\frac{1}{8}(D_{\alpha}F_iD^{\alpha}F_i)(F_jF_j)-\frac{1}{4}(D_{\alpha}F_iF_j)(D^{\alpha}F_iF_j)\biggr)\,. \nn 
\end{eqnarray}

\subsubsection{Eight Dimensions}
\begin{eqnarray}
&&C^{\oneloop,\,D=8}_{(R,R,R,R)}=-\frac{1}{\epsilon}\frac{1}{(4\pi)^4}\frac{1}{11520} \nn \\
&&\hspace{2.4cm}\null 
\times\left[16(238+D_s)\left(\frac{1}{2}R_{\mu\nu\lambda\gamma}R^{\nu\rho\gamma\delta}R_{\rho\sigma\delta\kappa}R^{\sigma\mu\kappa\lambda}+R_{\mu\nu\lambda\gamma}R^{\nu\rho}_{\hphantom{\nu\rho}\delta\kappa}R^{\hphantom{\rho\sigma}\gamma\delta}_{\rho\sigma}R^{\sigma\mu\kappa\lambda}\right)\right. \nn \\
&&\hspace{2.6cm}\null
+5(50-D_s)\left(\frac{1}{2}(R_{\mu\nu\rho\sigma}R^{\mu\nu\rho\sigma})^2+R_{\mu\nu\lambda\gamma}R^{\mu\nu}_{\hphantom{\mu\nu}\delta\kappa}R_{\rho\sigma}^{\hphantom{\rho\sigma}\lambda\gamma}R^{\rho\sigma\delta\kappa}\right) \nn \\
&&\hspace{2.6cm}\left.\null
 -16(122-D_s)\left(R_{\mu\nu\lambda\gamma}R^{\nu\rho\lambda\gamma}R_{\rho\sigma\delta\kappa}R^{\sigma\mu\delta\kappa}+\frac{1}{2}R_{\mu\nu\lambda\gamma}R^{\nu\rho}_{\hphantom{\nu\rho}\delta\kappa}R^{\hphantom{\rho\sigma}\lambda\gamma}_{\rho\sigma}R^{\sigma\mu\delta\kappa}\right)\right]\,, \nn \\
&&C^{\oneloop,\,D=8}_{(R,R,F,F)}=\frac{1}{\epsilon}\frac{1}{(4\pi)^4}\left(\frac{\kappa}{2}\right)^2\frac{1}{90} \nn \\
&&\hspace{3cm}\null 
\times\bigg[(D_s-32)\left((RRD_{\alpha}FD^{\alpha}F)+\frac{1}{2}(RD_{\alpha}FRD^{\alpha}F)\right. \nn \\
&&\hspace{6.5cm}\null 
-\left.\frac{1}{8}(RR)(D_{\alpha}FD^{\alpha}F)-\frac{1}{4}(RD_{\alpha}F)(RD^{\alpha}F)\right) \nn \\
&&\hspace{3.5cm}\null 
+60\left((R_{\mu\nu}R^{\mu}_{\hphantom{\mu}\rho}D^{\nu}FD^{\rho}F)+(R_{\mu\nu}R^{\mu}_{\hphantom{\mu}\rho}D^{\rho}FD^{\nu}F)\vphantom{\frac{1}{4}}\right. \nn \\
&&\hspace{5.5cm}\null
+(R_{\mu\nu}D^{\nu}FR^{\mu}_{\hphantom{\mu}\rho}D^{\rho}F)-\frac{1}{4}(R_{\mu\nu}R^{\mu}_{\hphantom{\mu}\rho})(D^{\nu}FD^{\rho}F) \nn \\
&&\hspace{5.5cm}\left.\null
-\frac{1}{4}(R_{\mu\nu}D^{\nu}F)(R^{\mu}_{\hphantom{\mu}\rho}D^{\rho}F)-\frac{1}{4}(R_{\mu\nu}D_{\rho}F)(R^{\mu\rho}D^{\nu}F)\right)\bigg]\,, \nn\\
&&C^{\oneloop,\,D=8}_{(F,F,F,F)}=-\frac{1}{\epsilon}\frac{1}{(4\pi)^4}\left(\frac{\kappa}{2}\right)^4\frac{D_s+18}{60}\left((D_{\alpha}FD^{\alpha}FD_{\beta}FD^{\beta}F)+\frac{1}{2}(D_{\alpha}FD_{\beta}FD^{\alpha}FD^{\beta}F)\right. \nn \\
&&\hspace{7cm}\left.\null
-\frac{1}{8}(D_{\alpha}FD^{\alpha}F)^2-\frac{1}{4}(D_{\alpha}FD_{\beta}F)(D^{\alpha}FD^{\beta}F)\right)\,, \nn \\
&&C^{\oneloop,\,D=8}_{(F_i,F_i,F_j,F_j)}=-\frac{1}{\epsilon}\frac{1}{(4\pi)^4}\left(\frac{\kappa}{2}\right)^4\frac{1}{60} \nn \\
&&\hspace{3.3cm}\null
\times\bigg[(D_s-22)\left((D_{\alpha}F_iD^{\alpha}F_iD_{\beta}F_jD^{\beta}F_j)+\frac{1}{2}(D_{\alpha}F_iD_{\beta}F_jD^{\alpha}F_iD^{\beta}F_j)\right. \nn \\
&&\hspace{5.3cm}\left.\null
-\frac{1}{8}(D_{\alpha}F_iD^{\alpha}F_i)(D_{\beta}F_jD^{\beta}F_j)-\frac{1}{4}(D_{\alpha}F_iD_{\beta}F_j)(D^{\alpha}F_iD^{\beta}F_j)\right) \nn \\
&&\hspace{4.3cm}\null 
+20\left((D_{\alpha}F_iD_{\beta}F_iD^{\alpha}F_jD^{\beta}F_j)+(D_{\alpha}F_iD^{\alpha}F_jD_{\beta}F_jD^{\beta}F_i)\vphantom{\frac{1}{4}}\right. \nn \\
&&\hspace{5.3cm}\null
+(D_{\alpha}F_iD_{\beta}F_jD^{\beta}F_iD^{\alpha}F_j)-\frac{1}{4}(D_{\alpha}F_iD_{\beta}F_i)(D^{\alpha}F_jD^{\beta}F_j) \nn \\
&&\hspace{5.3cm}\left.\null
-\frac{1}{4}(D_{\alpha}F_iD^{\alpha}F_j)^2-\frac{1}{4}(D_{\alpha}F_iD_{\beta}F_j)(D^{\beta}F_iD^{\alpha}F_j)\right)\bigg]\,.
\end{eqnarray}

\subsection{Two Loops}

\subsubsection{Four Dimensions}

\begin{eqnarray}
&&C^{\twoloop,\,D=4}_{(R,R,R,R)}=0\,, \nn\\
&&C^{\twoloop,\,D=4}_{(R,R,F,F)}=0\,, \nn \\
&&C^{\twoloop,\,D=4}_{(F,F,F,F)}=0\,, \\
&&C^{\twoloop,\,D=4}_{(F_i,F_i,F_j,F_j)}=\frac{1}{\epsilon^2}\frac{1}{(4\pi)^4}
\Bigl(\frac{\kappa}{2}\Bigr)^6
\frac{(D_s-2)^2}{4}\left((D_{\alpha}F_iD^{\alpha}F_iF_jF_j)+\frac{1}{2}(D_{\alpha}F_iF_jD^{\alpha}F_iF_j)\right. \nn \\
&&\hspace{6.5cm}-\left.\frac{1}{8}(D_{\alpha}F_iD^{\alpha}F_i)(F_jF_j)-\frac{1}{4}(D_{\alpha}F_iF_j)(D^{\alpha}F_iF_j)\right)\,. \nn
\end{eqnarray}

\subsubsection{Five Dimensions}

\begin{eqnarray}
&&C^{\twoloop,\,D=5}_{(R,R,R,R)}=0\,, \nn \\
&&C^{\twoloop,\,D=5}_{(R,R,F,F)}=-\frac{1}{\epsilon}\frac{1}{(4\pi)^5}
\Bigl(\frac{\kappa}{2}\Bigr)^4\frac{(10-D_s)\pi}{3}
\left((RRD_{\alpha}FD^{\alpha}F)+\frac{1}{2}(RD_{\alpha}FRD^{\alpha}F)\right. \nn \\
&&\hspace{7.2cm}\null
-\left.\frac{1}{8}(RR)(D_{\alpha}FD^{\alpha}F)-\frac{1}{4}(RD_{\alpha}F)(RD^{\alpha}F))\right)\,, \nn \\
&&C^{\twoloop,\,D=5}_{(F,F,F,F)}=\frac{1}{\epsilon}\frac{1}{(4\pi)^5}\Bigl(\frac{\kappa}{2}\Bigr)^6\frac{2(10-D_s)\pi}{3}\left((D_{\alpha}FD^{\alpha}FD_{\beta}FD^{\beta}F)\vphantom{\frac{1}{4}}\right. \nn \\
&&\hspace{2.5cm}\left.\null 
+\frac{1}{2}(D_{\alpha}FD_{\beta}FD^{\alpha}FD^{\beta}F)\vphantom{\frac{1}{4}}-\frac{1}{8}(D_{\alpha}FD^{\alpha}F)^2-\frac{1}{4}(D_{\alpha}FD_{\beta}F)(D^{\alpha}FD^{\beta}F)\right)\,, \nn \\
&&C^{\twoloop,\,D=5}_{(F_i,F_i,F_j,F_j)}=\frac{1}{\epsilon}\frac{1}{(4\pi)^5}\Bigl(\frac{\kappa}{2}\Bigr)^6\frac{2(10-D_s)\pi}{3} \nn \\
&&\hspace{3.5cm}\null
\times\left(\frac{3}{2}(D_{\alpha}F_iD^{\alpha}F_iD_{\beta}F_jD^{\beta}F_j)+\frac{3}{4}(D_{\alpha}F_iD_{\beta}F_jD^{\alpha}F_iD^{\beta}F_j)\right. \\
&&\hspace{4.5cm}\null 
+(D_{\alpha\beta}F_iF_iD^{\alpha}F_jD^{\beta}F_j)+\frac{1}{2}(D_{\alpha\beta}F_iD^{\alpha}F_jF_iD^{\beta}F_j) \nn \\
&&\hspace{4.5cm}\null
-\frac{3}{16}(D_{\alpha}F_iD^{\alpha}F_i)(D_{\beta}F_jD^{\beta}F_j)-\frac{3}{8}(D_{\alpha}F_iD_{\beta}F_j)(D^{\alpha}F_iD^{\beta}F_j) \nn \\
&&\hspace{4.5cm}\left.\null
-\frac{1}{8}(D_{\alpha\beta}F_iF_i)(D^{\alpha}F_jD^{\beta}F_j)-\frac{1}{4}(D_{\alpha\beta}F_iD^{\alpha}F_j)(F_iD^{\beta}F_j)\right)\,. \nn
\end{eqnarray}

\subsubsection{Six Dimensions}

\begin{flalign}
&\hspace{.5cm}C^{\twoloop,\,D=6}_{(R,R,R,R)}=-\frac{1}{(4\pi)^6}
 \Bigl(\frac{\kappa}{2}\Bigr)^2& \notag \\
&\hspace{2.5cm}\null
\times\bigg[\left(\frac{(D_s-6)(26-D_s)}{576\epsilon^2}-\frac{734-19D_s}{864\epsilon}\right)& \notag \\
&\hspace{3.2cm}\null
\times\left(D_{\alpha}R_{\mu\nu\lambda\gamma}D^{\alpha}R^{\mu\nu\gamma\delta}R_{\rho\sigma\delta\kappa}R^{\rho\sigma\kappa\lambda}+\frac{1}{2}D_{\alpha}R_{\mu\nu\lambda\gamma}R_{\rho\sigma}^{\hphantom{\rho\sigma}\gamma\delta}D^{\alpha}R^{\mu\nu}_{\hphantom{\mu\nu}\delta\kappa}R^{\rho\sigma\kappa\lambda}\right.& \notag \\
&\hspace{7cm}\left.\null
-\frac{1}{8}(D_{\alpha}R_{\mu\nu\lambda\gamma})^2(R_{\rho\sigma\delta\kappa})^2-\frac{1}{4}(D_{\alpha}R_{\mu\nu\lambda\gamma}R_{\rho\sigma}^{\hphantom{\rho\sigma}\lambda\gamma})^2\right)& \notag \\
&\hspace{2.7cm}\null 
-\frac{26-D_s}{18\epsilon}& \notag \\
&\hspace{3.2cm}\null 
\times\left(D_{\alpha}R_{\mu\nu\lambda\gamma}D^{\alpha}R_{\rho\sigma}^{\hphantom{\rho\sigma}\gamma\delta}R^{\nu\rho}_{\hphantom{\nu\rho}\delta\kappa}R^{\sigma\mu\kappa\lambda}+\frac{1}{2}D_{\alpha}R_{\mu\nu\lambda\gamma}R^{\nu\rho\gamma\delta}D^{\alpha}R_{\rho\sigma\delta\kappa}R^{\sigma\mu\kappa\lambda}\right.& \notag \\
&\hspace{3.5cm}\left.\null
-\frac{1}{8}D_{\alpha}R_{\mu\nu\lambda\gamma}R_{\rho\sigma}^{\hphantom{\rho\sigma}\gamma\delta}D^{\alpha}R^{\mu\nu}_{\hphantom{\mu\nu}\delta\kappa}R^{\rho\sigma\kappa\lambda}-\frac{1}{4}D_{\alpha}R_{\mu\nu\lambda\gamma}R^{\mu\nu\gamma\delta}D^{\alpha}R_{\rho\sigma\delta\kappa}R^{\rho\sigma\kappa\lambda}\right)\bigg]\,,& \notag \\
&\hspace{.5cm}C^{\twoloop,\,D=6}_{(R,R,F,F)}=-\frac{1}{(4\pi)^6}\Bigl(\frac{\kappa}{2}\Bigr)^4\frac{1}{144}& \notag \\
&\hspace{2.8cm}\null
\times\bigg[\left(\frac{(26-D_s)(27-2D_s)}{\epsilon^2}+\frac{25(D_s+22)}{6\epsilon}\right)& \notag \\
&\hspace{3.5cm}\null 
\times\left((D_{\alpha}RD^{\alpha}RD_{\beta}FD^{\beta}F)+\frac{1}{2}(D_{\alpha}RD_{\beta}FD^{\alpha}RD^{\beta}F)\right.& \notag \\
&\hspace{4cm}\left.\null 
-\frac{1}{8}(D_{\alpha}RD^{\alpha}R)(D_{\beta}FD^{\beta}F)-\frac{1}{4}(D_{\alpha}RD_{\beta}F)(D^{\alpha}RD^{\beta}F)\right)& \notag \\
&\hspace{3cm}\null
+\left(\frac{10(26-D_s)}{\epsilon^2}-\frac{734-19D_s}{3\epsilon}\right)& \notag \\
&\hspace{3.5cm}\times\left((D_{\alpha}R_{\mu\nu}D^{\alpha}R^{\mu}_{\hphantom{\mu}\rho}D^{\nu}FD^{\rho}F)+(D_{\alpha}R_{\mu\nu}D^{\alpha}R^{\mu}_{\hphantom{\mu}\rho}D^{\rho}FD^{\nu}F)\vphantom{\frac{1}{4}}\right.& \notag \\
&\hspace{4cm}\null 
+(D_{\alpha}R_{\mu\nu}D^{\nu}FD^{\alpha}R^{\mu}_{\hphantom{\mu}\rho}D^{\rho}F)-\frac{1}{4}(D_{\alpha}R_{\mu\nu}D^{\alpha}R^{\mu}_{\hphantom{\mu}\rho})(D^{\nu}FD^{\rho}F)& \notag \\
&\hspace{4cm}\left.\null
-\frac{1}{4}(D_{\alpha}R_{\mu\nu}D^{\nu}F)(D^{\alpha}R^{\mu}_{\hphantom{\mu}\rho}D^{\rho}F)-\frac{1}{4}(D_{\alpha}R_{\mu\nu}D_{\rho}F)(D^{\alpha}R^{\mu\rho}D^{\nu}F)\right)& \notag \\
&\hspace{3cm}\null
+\left(\frac{16(26-D_s)}{\epsilon^2}-\frac{8(D_s+22)}{3\epsilon}\right)& \notag \\
&\hspace{3.5cm}\null
\times\left((R_{\mu\nu}D_{\alpha}R^{\mu}_{\hphantom{\mu}\rho}D^{\alpha\nu}FD^{\rho}F)+(R_{\mu\nu}D_{\alpha}R^{\mu}_{\hphantom{\mu}\rho}D^{\rho}FD^{\alpha\nu}F)\vphantom{\frac{1}{4}}\right.& \notag \\
&\hspace{4cm}\null
+(R_{\mu\nu}D_{\alpha}^{\hphantom{\alpha}\nu}FD^{\alpha}R^{\mu}_{\hphantom{\mu}\rho}D^{\rho}F)-\frac{1}{4}(R_{\mu\nu}D_{\alpha}R^{\mu}_{\hphantom{\mu}\rho})(D^{\alpha\nu}FD^{\rho}F)& \notag \\
&\hspace{4cm}\left.\null
-\frac{1}{4}(R_{\mu\nu}D_{\alpha}^{\hphantom{\alpha}\nu}F)(D^{\alpha}R^{\mu}_{\hphantom{\mu}\rho}D^{\rho}F)-\frac{1}{4}(R_{\mu\nu}D_{\rho}F)(D_{\alpha}R^{\mu\rho}D^{\alpha\nu}F)\right)\bigg]\,,& \notag \displaybreak \\
&\hspace{.5cm}C^{\twoloop,\,D=6}_{(F,F,F,F)}=-\frac{1}{(4\pi)^6}\Bigl(\frac{\kappa}{2}\Bigr)^6\frac{1}{24}\left(\frac{(D_s-6)(26-D_s)}{\epsilon^2}-\frac{13D_s+142}{3\epsilon}\right)& \notag \\
&\hspace{5.5cm}\null
\times\left((D_{\alpha\beta}FD^{\alpha}_{\hphantom{\alpha}\gamma}FD^{\beta\gamma}FF)-\frac{1}{4}(D_{\alpha\beta}FD^{\alpha}_{\hphantom{\alpha}\gamma}F)(D^{\beta\gamma}FF)\right)\,,& \notag \\
&\hspace{.5cm}C^{\twoloop,\,D=6}_{(F_i,F_i,F_j,F_j)}=\frac{1}{(4\pi)^6}\Bigl(\frac{\kappa}{2}\Bigr)^6\frac{1}{36}& \notag \\
&\hspace{3.2cm}\null
\times\bigg[\left(\frac{(26-D_s)(16-D_s)}{\epsilon^2}-\frac{2(218-13D_s)}{3\epsilon}\right)& \notag \\
&\hspace{3.9cm}\null
\times\left((D_{\alpha\beta}F_iD^{\alpha\beta}F_iD_{\gamma}F_jD^{\gamma}F_j)+\frac{1}{2}(D_{\alpha\beta}F_iD_{\gamma}F_jD^{\alpha\beta}F_iD^{\gamma}F_j)\right.& \notag \\
&\hspace{4.5cm}\left.\null
-\frac{1}{8}(D_{\alpha\beta}F_iD^{\alpha\beta}F_i)(D_{\gamma}F_jD^{\gamma}F_j)-\frac{1}{4}(D_{\alpha\beta}F_iD_{\gamma}F_j)(D^{\alpha\beta}F_iD^{\gamma}F_j)\right)& \notag \\
&\hspace{3.83cm}\null
-\left(\frac{10(26-D_s)}{\epsilon^2}-\frac{578-13D_s}{3\epsilon}\right)& \notag \\
&\hspace{3.9cm}\null 
\times\left((D_{\alpha\beta}F_iD^{\alpha}_{\hphantom{\alpha}\gamma}F_iD^{\beta\gamma}F_jF_j)+\frac{1}{2}(D_{\alpha\beta}F_iD^{\beta}_{\hphantom{\beta}\gamma}F_jD^{\alpha\gamma}F_iF_j)\right.& \notag \\
&\hspace{4.7cm}\left.\null
-\frac{1}{8}(D_{\alpha\beta}F_iD^{\alpha}_{\hphantom{\alpha}\gamma}F_i)(D^{\beta\gamma}F_jF_j)-\frac{1}{4}(D_{\alpha\beta}F_iD^{\beta}_{\hphantom{\beta}\gamma}F_j)(D^{\alpha\gamma}F_iF_j)\right)\bigg]\,.&
\end{flalign}
%

%%%%%%%%%%%%%%%%%%%%%%%%%%%%%%%%%%%%%%%%%%%%%%%%%%%


\begin{thebibliography}{99}

%+% 1 ref
\bibitem{Supergravity}
 D.~Z.~Freedman, P.~van Nieuwenhuizen and S.~Ferrara,
%``Progress Toward a Theory of Supergravity,''
Phys.\ Rev.\ D {\bf 13}, 3214 (1976);\\
%%CITATION = PHRVA,D13,3214;%%
%
S.~Deser and B.~Zumino,
%``Consistent Supergravity,''
Phys.\ Lett.\ B {\bf 62}, 335 (1976).
%%CITATION = PHLTA,B62,335;%%

%+% 1 ref
\bibitem{HoweStelleReview}
P.~S.~Howe and K.~S.~Stelle,
%``The Ultraviolet Properties Of Supersymmetric Field Theories,''
Int.\ J.\ Mod.\ Phys.\ A {\bf 4}, 1871 (1989).
%%CITATION = IMPAE,A4,1871;%%

%+% 1 ref
\bibitem{UnitarityMethod}
Z.~Bern, L.~J.~Dixon, D.~C.~Dunbar and D.~A.~Kosower,
%``One loop n point gauge-theory amplitudes, unitarity and collinear limits,''
Nucl.\ Phys.\ B {\bf 425}, 217 (1994)
[hep-ph/9403226];
%%CITATION = HEP-PH/9403226;%%
%
%  Z.~Bern, L.~J.~Dixon, D.~C.~Dunbar and D.~A.~Kosower,
%``Fusing gauge theory tree amplitudes into loop amplitudes,''
Nucl.\ Phys.\ B {\bf 435}, 59 (1995)
[hep-ph/9409265].
%%CITATION = HEP-PH/9409265;%%

%+% 6 refs
\bibitem{BCJ}
Z.~Bern, J.~J.~M.~Carrasco and H.~Johansson,
%``New Relations for Gauge-Theory Amplitudes,''
Phys.\ Rev.\ D {\bf 78}, 085011 (2008)
[arXiv:0805.3993 [hep-ph]].
%%CITATION = ARXIV:0805.3993;%%

%+% 7 refs
\bibitem{BCJLoop}
Z.~Bern, J.~J.~M.~Carrasco and H.~Johansson,
%``Perturbative Quantum Gravity as a Double Copy of Gauge Theory,''
Phys.\ Rev.\ Lett.\  {\bf 105}, 061602 (2010)
[arXiv:1004.0476 [hep-th]].
%%CITATION = ARXIV:1004.0476;%%

%+% 1 ref
\bibitem{GravityThree}
Z.~Bern, J.~J.~M.~Carrasco, L.~J.~Dixon, H.~Johansson, D.~A.~Kosower 
and R.~Roiban,
%``Three-Loop Superfiniteness of N=8 Supergravity,''
Phys.\ Rev.\ Lett.\  {\bf 98}, 161303 (2007)
[arXiv:hep-th/0702112];\\
%%CITATION = PRLTA,98,161303;%%
%
Z.~Bern, J.~J.~M.~Carrasco, L.~J.~Dixon, H.~Johansson and R.~Roiban,
%``Manifest Ultraviolet Behavior for the Three-Loop Four-Point Amplitude of
%N=8 Supergravity,''
Phys.\ Rev.\  D {\bf 78}, 105019 (2008)
[arXiv:0808.4112 [hep-th]].
%%CITATION = PHRVA,D78,105019;%%

%+% 2 refs
\bibitem{GravityFour}
Z.~Bern, J.~J.~M.~Carrasco, L.~J.~Dixon, H.~Johansson and R.~Roiban,
%``The Ultraviolet Behavior of N=8 Supergravity at Four Loops,''
Phys.\ Rev.\ Lett.\  {\bf 103}, 081301 (2009)
[arXiv:0905.2326 [hep-th]].
%%CITATION = PRLTA,103,081301;%%

%+% 1 ref
\bibitem{N8Sugra}
E.~Cremmer and B.~Julia,
%``The SO(8) Supergravity,''
Nucl.\ Phys.\ B {\bf 159}, 141 (1979).
%%CITATION = NUPHA,B159,141;%%

%+% 1 ref
\bibitem{SevenLoopGravity}
M.~B.~Green, J.~G.~Russo and P.~Vanhove,
%``String-theory dualities and supergravity divergences,''
JHEP {\bf 1006}, 075 (2010)
[arXiv:1002.3805 [hep-th]];\\
%
J.~Bj\"{o}rnsson and M.~B.~Green,
%``5 loops in 24/5 dimensions,''
JHEP {\bf 1008}, 132 (2010)
[arXiv:1004.2692 [hep-th]];\\
%%CITATION = JHEPA,1008,132;%%
%\
G.~Bossard, C.~Hillmann and H.~Nicolai,
%``E7(7) symmetry in perturbatively quantised N=8 supergravity,''
JHEP {\bf 1012}, 052 (2010)
[arXiv:1007.5472 [hep-th]];\\
%%CITATION = ARXIV:1007.5472;%%
%
J.~Bj\"{o}rnsson,
  %``Multi-loop amplitudes in maximally supersymmetric pure spinor field theory,''
  JHEP {\bf 1101}, 002 (2011)
  [arXiv:1009.5906 [hep-th]]; \\
  %%CITATION = ARXIV:1009.5906;%%
%
G.~Bossard, P.~S.~Howe and K.~S.~Stelle,
%``On duality symmetries of supergravity invariants,''
JHEP {\bf 1101}, 020 (2011)
[arXiv:1009.0743 [hep-th]];\\
%%CITATION = ARXIV:1009.0743;%%
%
N.~Beisert, H.~Elvang, D.~Z.~Freedman, M.~Kiermaier, A.~Morales
and S.~Stieberger,
%``E7(7) constraints on counterterms in N=8 supergravity,''
Phys.\ Lett.\ B {\bf 694}, 265 (2010)
[arXiv:1009.1643 [hep-th]].
%%CITATION = ARXIV:1009.1643;%%

%+% 3 refs
\bibitem{VanishingVolume}
G.~Bossard, P.~S.~Howe, K.~S.~Stelle and P.~Vanhove,
%``The vanishing volume of D=4 superspace,''
Class.\ Quant.\ Grav.\  {\bf 28}, 215005 (2011)
[arXiv:1105.6087 [hep-th]].
%%CITATION = ARXIV:1105.6087;%%

%+% 1 ref
\bibitem{N4Sugra}
E.~Cremmer, J.~Scherk and S.~Ferrara,
%``SU(4) Invariant Supergravity Theory,''
Phys.\ Lett.\ B {\bf 74}, 61 (1978).
%%CITATION = PHLTA,B74,61;%%

%+% 8 refs
\bibitem{N4gravThreeLoops}
Z.~Bern, S.~Davies, T.~Dennen and Y.-t.~Huang,
%``Absence of Three-Loop Four-Point Divergences in N=4 Supergravity,''
Phys.\ Rev.\ Lett.\  {\bf 108}, 201301 (2012)
[arXiv:1202.3423 [hep-th]].
%%CITATION = ARXIV:1202.3423;%%

%+% 25 refs
\bibitem{HalfMax5D}
Z.~Bern, S.~Davies, T.~Dennen and Y.-t.~Huang,
%``Ultraviolet Cancellations in Half-Maximal Supergravity as a 
% Consequence of the Double-Copy Structure,''
Phys.\ Rev.\ D {\bf 86}, 105014 (2012)
[arXiv:1209.2472 [hep-th]].
%%CITATION = ARXIV:1209.2472;%%

%+% 2 refs
\bibitem{VanhoveN4}
P.~Tourkine and P.~Vanhove,
%``An R^4 non-renormalisation theorem in N=4 supergravity,''
Class.\ Quant.\ Grav.\  {\bf 29}, 115006 (2012)
[arXiv:1202.3692 [hep-th]].
%%CITATION = ARXIV:1202.3692;%%

%+% 5 refs
\bibitem{BossardHoweStelle5D}
G.~Bossard, P.~S.~Howe and K.~S.~Stelle,
%``Anomalies and divergences in N=4 supergravity,''
Phys.\ Lett.\ B {\bf 719}, 424 (2013)
[arXiv:1212.0841 [hep-th]].
%%CITATION = ARXIV:1212.0841;%%

%+% 11 refs
\bibitem{BHSNew}
G.~Bossard, P.~S.~Howe and K.~S.~Stelle,
%``Invariants and divergences in half-maximal supergravity theories,''
arXiv:1304.7753 [hep-th].
%%CITATION = ARXIV:1304.7753;%%

%+% 1 ref
\bibitem{RenataSergioN4}
  S.~Ferrara, R.~Kallosh and A.~Van Proeyen,
  %``Conjecture on Hidden Superconformal Symmetry of N=4 Supergravity,''
  Phys.\ Rev.\ D {\bf 87}, 025004 (2013)
  [arXiv:1209.0418 [hep-th]].
  %%CITATION = ARXIV:1209.0418;%%
  %8 citations counted in INSPIRE as of 21 May 2013

%+% 1 ref
\bibitem{KalloshN4}
R.~Kallosh,
%``On Absence of 3-loop Divergence in N=4 Supergravity,''
Phys.\ Rev.\ D {\bf 85}, 081702 (2012)
[arXiv:1202.4690 [hep-th]].
%%CITATION = ARXIV:1202.4690;%%

%+% 3 refs
\bibitem{deRoo}
M.~de Roo,
%``Matter Coupling in N=4 Supergravity,''
Nucl.\ Phys.\ B {\bf 255}, 515 (1985).
%%CITATION = NUPHA,B255,515;%%

%+% 2 refs
\bibitem{Awada}
M.~Awada and P.~K.~Townsend,
%``N=4 Maxwell-Einstein Supergravity In Five-dimensions And 
%its SU(2) Gauging,''
Nucl.\ Phys.\ B {\bf 255}, 617 (1985).
%%CITATION = NUPHA,B255,617;%%

%+% 2 refs
\bibitem{MarcusAnomaly}
N.~Marcus,
%``Composite Anomalies In Supergravity,''
Phys.\ Lett.\ B {\bf 157}, 383 (1985).
%%CITATION = PHLTA,B157,383;%%

%+% 1 ref
\bibitem{SugraMatter}
S.~Ferrara, J.~Scherk and B.~Zumino,
%``Supergravity and Local Extended Supersymmetry,''
Phys.\ Lett.\ B {\bf 66}, 35 (1977);\\
%%CITATION = PHLTA,B66,35;%%
%
S.~Deser, J.~H.~Kay and K.~S.~Stelle,
%``Renormalizability Properties of Supergravity,''
Phys.\ Rev.\ Lett.\  {\bf 38}, 527 (1977);\\
%%CITATION = PRLTA,38,527;%%
%
K.~S.~Stelle and P.~C.~West,
%``Relation Between Vector And Scalar Multiplets And Gauge Invariance In Supergravity,''
Nucl.\ Phys.\ B {\bf 145}, 175 (1978).
%%CITATION = NUPHA,B145,175;%%

%+% 4 refs
\bibitem{Fischler}
M.~Fischler,
%``Finiteness Calculations For O(4) Through O(8) Extended Supergravity And O(4) Supergravity Coupled To Selfdual O(4) Matter,''
Phys.\ Rev.\ D {\bf 20}, 396 (1979).
%%CITATION = PHRVA,D20,396;

%+% 4 refs
\bibitem{Arkady}
E.~S.~Fradkin and A.~A.~Tseytlin,
%``One Loop Infinities In Dimensionally Reduced Supergravities,''
Phys.\ Lett.\ B {\bf 137}, 357 (1984).
%%CITATION = PHLTA,B137,357;%%

%+% 2 refs
\bibitem{BRY}
Z.~Bern, J.~S.~Rozowsky and B.~Yan,
%``Two-loop four-gluon amplitudes in N = 4 super-Yang-Mills,''
Phys.\ Lett.\  B {\bf 401}, 273 (1997)
[arXiv:hep-ph/9702424].
%%CITATION = PHLTA,B401,273;%%

%+% 3 refs
\bibitem{BDDPR}
Z.~Bern, L.~J.~Dixon, D.~C.~Dunbar, M.~Perelstein and J.~S.~Rozowsky,
%``On the relationship between Yang-Mills theory and gravity and its
%implication for ultraviolet divergences,''
Nucl.\ Phys.\  B {\bf 530}, 401 (1998)
[arXiv:hep-th/9802162].
%%CITATION = NUPHA,B530,401;%%

%+% 2 refs
\bibitem{RaduMatter}
J.~J.~M.~Carrasco, M.~Chiodaroli, M.~G\"unaydin and R.~Roiban,
%``One-loop four-point amplitudes in pure and matter-coupled N <= 4 supergravity,''
JHEP {\bf 1303}, 056 (2013)
[arXiv:1212.1146 [hep-th]].
%%CITATION = ARXIV:1212.1146;%%

%+% 3 refs
\bibitem{MarcusSagnotti}
A.~A.~Vladimirov,
%``Method For Computing Renormalization Group Functions In Dimensional
%Renormalization Scheme,''
Theor.\ Math.\ Phys.\  {\bf 43}, 417 (1980)
[Teor.\ Mat.\ Fiz.\  {\bf 43}, 210 (1980)];\\
%%CITATION = TMFZA,43,210;%%
%
N.~Marcus and A.~Sagnotti,
%``A Simple Method For Calculating Counterterms,''
Nuovo Cim.\ A {\bf 87}, 1 (1985).
%%CITATION = NUCIA,A87,1;%%

%+% 1 ref
\bibitem{ck4l}
Z.~Bern, J.~J.~M.~Carrasco, L.~J.~Dixon, H.~Johansson and R.~Roiban,
%``Simplifying Multiloop Integrands and Ultraviolet Divergences of Gauge Theory and Gravity Amplitudes,''
Phys.\ Rev.\ D {\bf 85}, 105014 (2012)
[arXiv:1201.5366 [hep-th]].
%%CITATION = ARXIV:1201.5366;%%

%+% 1 ref
\bibitem{HenrikJJReview}
J.~J.~M.~Carrasco and H.~Johansson,
%``Generic multiloop methods and application to N=4 super-Yang-Mills,''
J.\ Phys.\ A A {\bf 44}, 454004 (2011)
[arXiv:1103.3298 [hep-th]].
%%CITATION = ARXIV:1103.3298;%%

%+% 7 refs
\bibitem{OneLoopN4}
Z.~Bern, C.~Boucher-Veronneau and H.~Johansson,
%``N >= 4 Supergravity Amplitudes from Gauge Theory at One Loop,''
Phys.\ Rev.\  D {\bf 84}, 105035 (2011)
[arXiv:1107.1935 [hep-th]].
%%CITATION = PHRVA,D84,105035;%%

%+% 5 refs
\bibitem{TwoLoopN4}
C.~Boucher-Veronneau and L.~J.~Dixon,
%``N >= 4 Supergravity Amplitudes from Gauge Theory at Two Loops,''
JHEP {\bf 1112}, 046 (2011) [arXiv:1110.1132 [hep-th]].
%%CITATION = JHEPA,1112,046;%%

%+% 1 ref
\bibitem{BCJSquare}
Z.~Bern, T.~Dennen, Y.-t.~Huang and M.~Kiermaier,
%``Gravity as the Square of Gauge Theory,''
Phys.\ Rev.\  D {\bf 82}, 065003 (2010)
[arXiv:1004.0693 [hep-th].
%%CITATION = PHRVA,D82,065003;%%

%+% 1 ref
\bibitem{KLT}
H.~Kawai, D.~C.~Lewellen and S.-H.~H.~Tye,
%``A Relation Between Tree Amplitudes Of Closed And Open Strings,
Nucl.\ Phys.\ B {\bf 269}, 1 (1986);\\
%%CITATION = NUPHA,B269,1;%%
%
Z.~Bern,
%``Perturbative quantum gravity and its relation to gauge theory,''     
Living Rev.\ Rel.\  {\bf 5}, 5 (2002)
[gr-qc/0206071].
%%CITATION = 00222,5,5;%%

%+% 2 refs
\bibitem{DixonMaltoniColor}
V.~Del Duca, L.~J.~Dixon and F.~Maltoni,
%``New color decompositions for gauge amplitudes at tree and loop level,''
Nucl.\ Phys.\ B {\bf 571}, 51 (2000)
[hep-ph/9910563].
%%CITATION = HEP-PH/9910563;%%

%+% 1 ref
\bibitem{ColorLoop}
Z.~Bern and D.~A.~Kosower,
%``Color decomposition of one loop amplitudes in gauge theories,''
Nucl.\ Phys.\ B {\bf 362}, 389 (1991).
%%CITATION = NUPHA,B362,389;%%

%+% 1 ref
\bibitem{GSB}
M.~B.~Green, J.~H.~Schwarz and L.~Brink,
%``N=4 Yang-Mills And N=8 Supergravity As Limits Of String Theories,''
Nucl.\ Phys.\  B {\bf 198}, 474 (1982).
%%CITATION = NUPHA,B198,474;%%

%+% 1 ref
\bibitem{SuperSum}
 Z.~Bern, J.~J.~M.~Carrasco, H.~Ita, H.~Johansson and R.~Roiban,
  %``On the Structure of Supersymmetric Sums in Multi-Loop Unitarity Cuts,''
  Phys.\ Rev.\ D {\bf 80}, 065029 (2009)
  [arXiv:0903.5348 [hep-th]].
  %%CITATION = ARXIV:0903.5348;%%
  %35 citations counted in INSPIRE as of 04 May 2013

%+% 1 ref
\bibitem{DunbarNorridge}
D.~C.~Dunbar and P.~S.~Norridge,
%``Calculation of graviton scattering amplitudes using string based methods,''
Nucl.\ Phys.\ B {\bf 433}, 181 (1995)
[hep-th/9408014];\\
%%CITATION = HEP-TH/9408014;%%
%
D.~C.~Dunbar, J.~H.~Ettle and W.~B.~Perkins,
%``Perturbative expansion of N <8 Supergravity,''
Phys.\ Rev.\ D {\bf 83}, 065015 (2011)
[arXiv:1011.5378 [hep-th]].
%%CITATION = ARXIV:1011.5378;%%

%+% 2 refs
\bibitem{Neq44np}
Z.~Bern, J.~J.~M.~Carrasco, L.~J.~Dixon, H.~Johansson and R.~Roiban,
%``The Complete Four-Loop Four-Point Amplitude in N=4 Super-Yang-Mills
%Theory,''
Phys.\ Rev.\  D {\bf 82}, 125040 (2010)
[arXiv:1008.3327 [hep-th]].
%%CITATION = PHRVA,D82,125040;%%

%+% 1 ref
\bibitem{Naculich}
S.~G.~Naculich,
%``All-loop group-theory constraints for color-ordered SU(N) gauge-theory 
% amplitudes,''
Phys.\ Lett.\ B {\bf 707}, 191 (2012)
[arXiv:1110.1859 [hep-th]].
%%CITATION = ARXIV:1110.1859;%%

%+% 1 ref
\bibitem{TwoLoopsYM}
Z.~Bern, A.~De Freitas and L.~J.~Dixon,
%``Two loop helicity amplitudes for gluon-gluon scattering in QCD and supersymmetric Yang-Mills theory,''
JHEP {\bf 0203}, 018 (2002)
[hep-ph/0201161].
%%CITATION = HEP-PH/0201161;%%

%+% 1 ref
\bibitem{SchnitzerIR}
S.~Weinberg,
%``Infrared photons and gravitons,''
Phys.\ Rev.\  {\bf 140}, B516 (1965);\\
%%CITATION = PHRVA,140,B516;%%
%
S.~G.~Naculich, H.~Nastase and H.~J.~Schnitzer,
%``Two-loop graviton scattering relation and IR behavior in N=8 supergravity,''
Nucl.\ Phys.\ B {\bf 805}, 40 (2008)
[arXiv:0805.2347 [hep-th]];\\
%%CITATION = ARXIV:0805.2347;%%
%
S.~G.~Naculich and H.~J.~Schnitzer,
%``Eikonal methods applied to gravitational scattering amplitudes,''
JHEP {\bf 1105}, 087 (2011)
[arXiv:1101.1524 [hep-th]];\\
%%CITATION = ARXIV:1101.1524;%%
%
R.~Akhoury, R.~Saotome and G.~Sterman,
%``Collinear and Soft Divergences in Perturbative Quantum Gravity,''
Phys.\ Rev.\ D {\bf 84}, 104040 (2011)
[arXiv:1109.0270 [hep-th]].
%%CITATION = ARXIV:1109.0270;%%

%+% 1 ref
\bibitem{Fire}
A.~V.~Smirnov,
%``Algorithm FIRE -- Feynman Integral REduction,''
JHEP {\bf 0810}, 107 (2008)
[arXiv:0807.3243 [hep-ph]].
%%CITATION = JHEPA,0810,107;%%

%+% 1 ref
\bibitem{MB}
M.~Czakon,
%``Automatized analytic continuation of Mellin-Barnes integrals,''
Comput.\ Phys.\ Commun.\  {\bf 175}, 559 (2006)
[hep-ph/0511200].
%%CITATION = HEP-PH/0511200;%%

%+% 1 ref
\bibitem{MellinBarnes}
V.~A.~Smirnov,
%``Analytical result for dimensionally regularized massless on shell double box,''
Phys.\ Lett.\ B {\bf 460}, 397 (1999)
[hep-ph/9905323];
%%CITATION = HEP-PH/9905323;%%

%+% 2 refs
\bibitem{FDH}
Z.~Bern and D.~A.~Kosower,
%``The Computation of loop amplitudes in gauge theories,''
Nucl.\ Phys.\  B {\bf 379}, 451 (1992);\\
%%CITATION = NUPHA,B379,451;%%
%
Z.~Bern, A.~De Freitas, L.~J.~Dixon and H.~L.~Wong,
%``Supersymmetric regularization, two-loop QCD amplitudes and coupling
%shifts,''
Phys.\ Rev.\  D {\bf 66}, 085002 (2002)
[arXiv:hep-ph/0202271].
%%CITATION = PHRVA,D66,085002;%%

%+% 1 ref
\bibitem{Dunbar}
D.~C.~Dunbar, B.~Julia, D.~Seminara and M.~Trigiante,
%``Counterterms in type I supergravities,''
JHEP {\bf 0001}, 046 (2000)
[hep-th/9911158].
%%CITATION = HEP-TH/9911158;%%

%+% 2 refs
\bibitem{RaduAnomaly}
J.~J.~M.~Carrasco, R.~Kallosh, R.~Roiban and A.~A.~Tseytlin,
%``On the U(1) duality anomaly and the S-matrix of N=4 supergravity,''
arXiv:1303.6219 [hep-th].
%%CITATION = ARXIV:1303.6219;%%

%+% 1 ref
\bibitem{HoweSuperspace}
P.~S.~Howe, H.~Nicolai and A.~Van Proeyen,
%``Auxiliary Fields And A Superspace Lagrangian For Linearized Ten-dimensional Supergravity,''
Phys.\ Lett.\ B {\bf 112}, 446 (1982).
%%CITATION = PHLTA,B112,446;%%

\end{thebibliography}
\end{document}